\documentclass[conference, compsoc]{IEEEtran}
\IEEEoverridecommandlockouts
\usepackage{url}

\usepackage{cite}
\usepackage{amsmath,amssymb,amsfonts}
\usepackage{algorithmic}
\usepackage{graphicx}
\usepackage{textcomp}
\usepackage[dvipsnames]{xcolor}

\usepackage{tikz}
\usepackage{xspace}

\usepackage[indentfirst=false, leftmargin=2.5ex,rightmargin=2.5ex,vskip=0.5ex]{quoting}

\usepackage[autostyle]{csquotes}
\MakeOuterQuote{"}

\usepackage{booktabs}
\usepackage{tabularx}
\usepackage{makecell}
\usepackage{csquotes}
\usepackage[most]{tcolorbox}

\usepackage[T1]{fontenc}
\usepackage{pifont}

\usepackage{listings}
\usepackage{inconsolata}
\usepackage{titlecaps}
\usepackage{listofitems}

\usepackage{hyperref}

\usepackage{comment}

\def\BibTeX{{\rm B\kern-.05em{\sc i\kern-.025em b}\kern-.08em
    T\kern-.1667em\lower.7ex\hbox{E}\kern-.125emX}}

\newboolean{showcomments}

\setboolean{showcomments}{true}

\ifthenelse{\boolean{showcomments}}
{
    
}

\newcommand\RED[1]{{\color{red}{#1}}}
\newcommand\GREEN[1]{{\color{OliveGreen}{#1}}}

\newcommand\angled[1]{{\texttt{<#1>}}}
\newcommand\myquoteinline[1]{{\textit{"#1"}}}
\newcommand\myquote[2]{{\begin{quoting}\textit{"#1" --- #2}\end{quoting}}\noindent}
\newcommand\myparagraph[1]{\noindent\underline{\bf {#1}:}}
\newcommand\myparagraphnew[1]{\vspace{0.25em}\noindent{\bf {#1}:}}

\newtcolorbox{mybox}[1][]{colframe=white,colback=white,
left=1.5pt,
right=1.5pt,
top=2pt,
bottom=1pt,
boxrule=0.3pt,
segmentation style={black, line width=0.5pt},
#1}

\newcounter{fcounter}
\newcommand{\finding}[1]{\refstepcounter{fcounter}
\vspace{0.5em}\noindent\fbox{%
    \parbox{0.95\linewidth}{%
  \vspace{0.3em}{\bf
  {Finding~\arabic{fcounter}~(\fnumber{\arabic{fcounter}})}~--} {#1}
\vspace{0.3em}
  }
}
\vspace{0.5em}

}
\newcommand\fnumber[1]{{$\mathcal{F}_{#1}$}}

\newcommand\tfootnotesymbol[1]{$^{\mathrm{#1}}$}
\newcommand\tfootnotecontent[2]{$^{\mathrm{#1}}$#2}

\readlist\participant{
    PAMP01\_7,
    PMP02\_7,
    PAMP03\_7,
    PAMP04\_7,
    SAMP05\_7,
    SAP06\_1,
    PMP07\_6,
    PAMP08\_7,
    PAMP09\_7,
    PMP10\_7,
    PAMP11\_7,
    PAMP12\_7,
    SMP13\_4,
    PAMP14\_7,
    SAMP15\_6,
    PAP16\_4,
    PMP17\_0,
    PAMP18\_5,
    SAMP19\_6,
    PAMP20\_7
}

\newcommand{\sast}{{SAST}\xspace}
\newcommand{\sasts}{{SASTs}\xspace}

\newcommand{\ie}{\textit{i.e.,}\xspace}
\newcommand{\eg}{\textit{e.g.,}\xspace}
\newcommand{\etc}{\textit{etc.}\xspace}
\newcommand{\etal}{\textit{et al.}\xspace}

\newcommand{\no}{{\RED{\ding{55}}}}
\newcommand{\ye}{\GREEN{\ding{51}}}

\newcommand{\countInterviews}{$20$}

\newcommand{\interviewSectionParticipants}{Participants, Projects, and Organizations}

\newcommand{\interviewSectionSecAndOrg}{Organization and Security}
\newcommand{\interviewSectionOrgOfSast}{Organizational Context of \sast}
\newcommand{\interviewSectionLimitExpectSast}{Expectations from \sast}
\newcommand{\interviewSectionImpactUnsoundSast}{Impact of Unsound/Flawed \sast}
\newcommand{\interviewSectionChallengeSolution}{Challenges and Improvements}

\newcounter{rqcounter}
\newcommand{\newrq}[2]{\noindent\refstepcounter{rqcounter}\textbf{RQ\arabic{rqcounter}:} {\em #2}\label{#1}}
\newcommand{\rqref}[1]{\textbf{RQ\ref{#1}}}

\excludecomment{bluetext}
\includecomment{blacktext}

\begin{bluetext}
    \newcommand\add[1]{{\color{blue} {#1}}}
    \newcommand\update[1]{{\color{blue} {#1}}}
\end{bluetext}

\begin{blacktext}
    \newcommand\add[1]{{{#1}}}
    \newcommand\update[1]{{{#1}}}
\end{blacktext}

\begin{document}

\title{"{\em False negative - that one is going to kill you}": \\  Understanding Industry Perspectives of Static Analysis based Security Testing}

\author{
  \IEEEauthorblockN{Amit Seal Ami\IEEEauthorrefmark{1},
Kevin Moran\IEEEauthorrefmark{2},
Denys Poshyvanyk\IEEEauthorrefmark{1}, and
Adwait Nadkarni\IEEEauthorrefmark{1}
}
\IEEEauthorblockA{\IEEEauthorrefmark{1}William \& Mary,
Williamsburg, VA, USA;\ {{aami@, denys@cs., nadkarni@cs.}wm.edu}}
\IEEEauthorblockA{\IEEEauthorrefmark{2}University of Central Florida, Orlando, FL, USA;\
kpmoran@ucf.edu
}}

\maketitle

\begin{abstract}
The demand for automated security analysis techniques, such as static analysis based security testing (SAST) tools continues to increase. 
To develop SASTs that are effectively leveraged by developers for finding vulnerabilities, researchers and tool designers must understand how developers perceive, select, and use SASTs, what they expect from the tools, whether they know of the limitations of the tools, and how they address those limitations.  
This paper describes a qualitative study that explores the assumptions, expectations, beliefs, and challenges experienced by developers who use SASTs.
We perform in-depth, semi-structured interviews with 20 practitioners who possess a diverse range of software development expertise, as well as a variety of unique security, product, and organizational backgrounds. We identify $17$ key findings that shed light on developer perceptions and desires related to SASTs, and also expose gaps in the status quo -- challenging long-held beliefs in SAST design priorities. Finally, we provide concrete future directions for researchers and practitioners rooted in an analysis of our findings.

\end{abstract}
    \vspace{-0.5em}
\section{Introduction}\label{sec:introduction}
\vspace{-0.5em}

Software security has gained continued international attention in recent years due to the increase of high-profile cyberattacks and exploits across the public sector. %
For example, incidents such as the SolarWinds Cyberattack
prompted the U.S. Government Accountability Office to elicit responses from both private and public sectors in 2021 to increase the effectiveness of security practices~\cite{solarwinds-response}.
Consequently, corporate and government entities alike are now increasingly emphasizing the security of software and services through a combination of $(1)$ new approaches (\eg Software Bill of Materials (SBOM)~\cite{certify-software-senate-2021}), $(2)$ adoption of security focused certifications of software (\eg Cyber Shield Act~\cite{cyber-shield-act}, IoT Compliance~\cite{ioxt23}), and $(3)$ improvement of existing approaches (\eg identifying and employing recommended types of automated software security testing~\cite{Hou21}). %
As a result, the existing multi-billion dollar industry of automated security analysis tools~\cite{gartner-application-testing-billion-dollars}, particularly Static Application Security Testing (SAST), have continued to proliferate to meet the increased security needs of organizations worldwide.
Further, such tools are now being incorporated into nearly every stage of the software development and maintenance lifecycle, from requirements engineering to fault localization and fixing (\eg the GitHub Code Scan initiative~\cite{github-code-scan}). %

However, SAST tools have been found to suffer from design and implementation flaws~\cite{ACK+22,AKM+21} that prevent them from detecting vulnerabilities that they claim to detect, or which can be expected for certain critical use cases they support (\eg compliance, audits).
Particularly, while SAST tools from industry, the open-source community, and academia have been found to support similar use cases, their design goals often differ dramatically~\cite{ACK+22}.
That is, tools may adopt a \textit{technique-centric} approach, wherein what they can detect is tied to the limitations of a set of chosen static analysis techniques, or, a {\em security-centric} approach, wherein the tool aims to use whichever static analysis techniques necessary to detect vulnerabilities falling under a specific security goal.
These different design ethos carry with them various trade-offs that impact the applicability, efficiency, and effectiveness of the security tools.
These trade-offs and their implications for cybersecurity in practice are currently poorly understood, at best.

In other words, we are increasingly heading towards a future where software developers will be depending more than ever on security focused program-analysis techniques for security assurance, compliance, and audits.
While we know of potential flaws in SASTs (as discussed previously), there exists a {\bf \em key gap} in prior research:
{\em the research community has only a limited understanding of how software developers perceive SASTs, what they expect from SASTs and believe in (particularly in terms of their ability to detect vulnerabilities), and how these perceptions and beliefs impact the adoption and use of SASTs in practice.}
Without addressing this gap through an understanding of the {\em practitioners' perspective}, we may not be able to develop SASTs that are truly effective in practice, \ie possess key properties that practitioners desire in order to improve software security, and moreover, will be unable to uncover gaps in what the practitioners (\ie users of the tools) expect or believe, versus what the tools actually provide, leading to a false sense of security.

\myparagraph{Contributions} This paper describes a qualitative study that investigates the assumptions, expectations, beliefs, and challenges experienced by practitioners who use program-analysis based security-assurance tools, specifically \sasts.
Our study is guided by 3 key research questions ({\bf RQ1} -- {\bf RQ3}), which we explore via in-depth interviews ($n=20$) with software developers, project managers, research engineers and practitioners, who together cover a broad range of security, product, and business contexts:

\newrq{1}{\update{How do practitioners at organizations, with different types of business and security needs, choose and depend on \sasts for ensuring security in their services/products?}}
Various factors may influence an organization's process for selecting a \sast tool, ranging from security or business needs (\eg compliance), brand reputation, or inclusion of safety-critical components. %
Thus, we are interested in exploring what individual practitioners and their organizations care about in terms of security, and how those needs affect the selection of SASTs, as well as their incorporation into their overall vulnerability detection processes.
We also seek to explore {\em how} SASTs are selected, \ie the subjective or objective processes involved in choosing a particular SAST.

\newrq{2}{What do practitioners know and believe about the limitations of \sasts, and what do they expect from them?}
While certain limitations, such as false positives, are relatively well-known, potential issues related to design and/or implementation flaws that result in security-specific false negatives are often unknown and unaccounted for in \sasts.
We are interested to understand the awareness, expectations, and beliefs of practitioners about such limitations, both known and unknown, of \sasts, particularly in terms of false positives and negatives.

\newrq{3}{How do practitioners navigate, address, or work around flaws of \sasts?} A \sast that does not detect vulnerabilities
may lead to vulnerabilities in otherwise security-assured software. %
We are interested in learning about practitioners' experiences regarding the impact of the flaws in \sasts (\eg product-related security incidents).
Furthermore, we are interested to know how practitioners balance the possibility of unsound \sasts that may make their product vulnerable, and the decision to release potentially vulnerable software.
Moreover, if practitioners do happen to find a flaw in a \sast, we seek to uncover their experiences in reporting the issues to the SASTs.
Finally, we seek to investigate the typical pain points that practitioners experience regarding SASTs, in order to understand the key properties they desire more than anything else.

    \section{Methodology}\label{sec:method}

To understand the potentially diverse perspectives of practitioners related to \sasts, we performed a two-phase study, composed of a survey, followed by in-depth, detailed interviews.
The purpose of the survey was to develop an initial understanding of the landscape, and more importantly, to guide the design of the interview protocol.
Therefore, this section (and the rest of the paper) focuses on the interviews and their qualitative analysis.
Moreover, the artifacts associated with the survey and the interview, including the informed consent forms, survey questionnaires, and the interview guide, are available in our online appendix~\cite{online-appendix}.

We now provide a brief summary of the survey protocol and results, which are further detailed in Appendix~\ref{app:online-survey} and Appendix~\ref{app:survey-results} respectively.
The interview protocol is described in Section~\ref{sec:interview-protocol}.

\subsection{Summary of the Survey Protocol and Results}
\label{sec:survey-protocol-results}
\vspace{-0.5em}
To understand how practitioners perceive and use SASTs and security tools more broadly, and how security is prioritized by individuals and organizations, we prepared an online survey questionnaire (the questionnaire is in the online appendix~\cite{online-appendix}) consisting of Likert-based questions, with optional open-ended responses to clarify their selected choice(s).
The survey protocol was approved by the Institutional Review Boards (IRBs) at the authors' universities.

\add{
We used two recruitment channels. First, we relied on {\em snowball sampling}~\cite{SnowballSamplinggoodman1961a}  from our professional networks primarily by sending invitation emails and requesting forwards to colleagues.
Second, we {\em emailed OSS developers} who had interacted with \sasts via CI/CD actions, \eg~GitHub Workflows, in open-source repositories that $(a)$~had at least one star or watcher, $(b)$ were not a fork, and $(c)$ used one of the top ten programming languages reported in GitHub Octoverse~\cite{octoverse2021}. We developed scripts that used GitHub Search APIs and crawled public repositories in Coverity Scan~\cite{coverity-scan} to find qualifying repositories. Next, we extracted email addresses from commits specific to CI/CD files (\eg {\small\texttt{.github/workflows/*.yml}}) that contained SAST names. Finally, we excluded any Github-assigned private emails and those that indicated no-reply.
\update{In the end, we contacted $1,918$ \update{OSS developers} exactly once via email.}
We discuss the ethical considerations in recruiting OSS developers in Appendix~\ref{app:ethical-oss-recruitment}.
}
\\

\vspace{-0.5em}

\update{\myparagraphnew{Survey Results Overview} We received $89$ responses from the survey, of which $39$ ($18$/$39$ responses from OSS developers) were complete and valid. Of these, 35/39 worked for organizations, whereas 2 were freelancers, and 2 chose not to indicate organizational status.
We made several observations that guided the design of our interview protocol based on these valid responses.}
\update{
\textit{First}, over $83$\% ($29/35$) participants (from organizations) stated that security was of "extreme" importance to them, with an additional $17$\% ($6/35$) stating it as "very" important.
In contrast, $63$\% ($22/35$) participants identified security as of "extreme" importance from their organization's perspective, with several ($8/35$ or $23$\%) considering it "very" important.
}
This tells us that security may be prioritized differently by the organization and the individual, and more importantly, that individuals may be willing to talk about these differences, which is critical for our interviews.
\textit{Second}, participants mostly relied on a combination of automated and manual security analyses, with some exceptions.
\textit{Finally}, almost all practitioners expressed that even in the case of flaws of \sasts, their applications would be moderately impacted at most, as  they rely on multiple tools and/or manual reviews.
These observations guided our protocol design for the interview phase, which we describe next.

\vspace{-0.5em}
\subsection{Interview Protocol}\label{sec:interview-protocol}
\vspace{-0.5em}
We drafted an interview protocol consisting of a semi-structured interview-guide~\cite{Ada15,HA05} structured by "laddering" questions~\cite{CRM+94}, a questioning strategy that is used to understand the relations between concepts in a domain and to explore the concepts in-depth.
The interview guide is designed to help understand the processes used to choose \sasts, how practitioners depend on \sasts for security assurance, their expectations about limitations of \sasts{}, and how they work around such limitations.
An abridged version of the interview guide is in Table~\ref{tbl:interview-questions} in the Appendix.

\vspace{-0.25em}
\subsubsection{Interview Recruitment}\label{sec:interview-recruitment}
We recruited $20$ interview participants through multiple recruitment channels, aiming for diversity in project, cultural background, experience and industry contexts.
Particularly, we recruited $(i)$ \update{$10$} participants from the survey and $(ii)$ $10$ separately through our professional network.
When recruiting from the survey-pool, we only invited participants who submitted reasonably valid responses and expressed interest in interview participation.
To recruit from our professional networks, we relied on snowball sampling~\cite{SnowballSamplinggoodman1961a}, i.e., emailed invitations to software engineers within our network, with details of the study as per protocol, and requested them to forward the invitation to colleagues experienced with \sasts. %

Overall, we recruited $20$ interview participants with diverse cultural background (\eg participants were from Asia, Europe, United Kingdom and North America, working in either local or international projects), industry contexts (\eg safety-critical, business-critical, research \& development, open-source \etc), experience ranging from entry level engineers to project managers, and security-context (\eg working towards compliance).
The anonymized details of participants are shown in Table~\ref{tbl:interview-participants}.  %

\vspace{-0.25em}
\subsubsection{Interview Protocol and Ethics}
\label{sec:ethics}

Similar to the survey, the final version of our interview guide and protocol was approved by our Institutional Review Boards (IRBs). %
Our consent form emphasized that no personally identifiable information would be collected, and responses will be anonymized even in the case of willfully shared private information, such as a participant's (or colleagues) name, associated previous or current organization, and product/client name(s).
Each participant would be interviewed for approximately an hour, and would receive a $\$50.00$ gift card or voucher in local currency.

Furthermore, as our interview-guide contained \textit{sensitive} questions \eg security enhancement vs meeting deadlines, or site-incidents due to flawed \sast,
we followed the Menlo Report Guidelines~\cite{KD12,DKB13} to refine our protocol and interview guide to avoid any potential harm, and reminded participants that they could withdraw/redact at any time, as further detailed in Section~\ref{sec:interview-procedure}.

\vspace{-0.25em}
\subsubsection{Interview Pilot \& Refinement}
We conducted pilot interviews, followed by an in-depth discussion, with three participants within our professional network to improve the interview guide.
Among these, one held a doctoral degree in computer science, with a focus in CyberSecurity, while the other two were pursuing a Ph.D. in Computer Science.

\vspace{-0.25em}
\subsubsection{Interviewing Procedure}
\label{sec:interview-procedure}
We conducted the interviews using either the lead-interviewer or lead-and-backup approaches, while following the semi-structured interview-guide.
The lead-and-backup approach ensured that each interviewer experienced conducting the interview with the guide.
While relevant questions from the guide were raised, it also allowed the lead interviewer to focus on listening and asking laddering/follow-up questions to the interviewee.
All the interviews were conducted virtually via Zoom.
We emailed the Informed Consent Form (and survey response when applicable) with IRB protocol references a day before the interview.
After the participant joined us in the online interview session, we reminded the participant \textit{before starting the interview} that $(a)$ the interview will be recorded, $(b)$ we will anonymize any sensitive information while transcribing, $(c)$ recorded audio will be destroyed after transcribing, $(d)$ they can redact anything they said at any point of the interview and/or can email us about it, and $(e)$ they have the option to stop the interview at any point.
The median, effective duration of the interviews, \ie excluding the intro, briefing and verbal consent, was $1$ hour $52$ seconds.

\vspace{-0.5em}
\subsection{Structure of the Interview Guide}\label{sec:interview-guide}
\vspace{-0.5em}

\begin{figure}[tbp]
	\centering
    \includegraphics[width=0.96\linewidth]{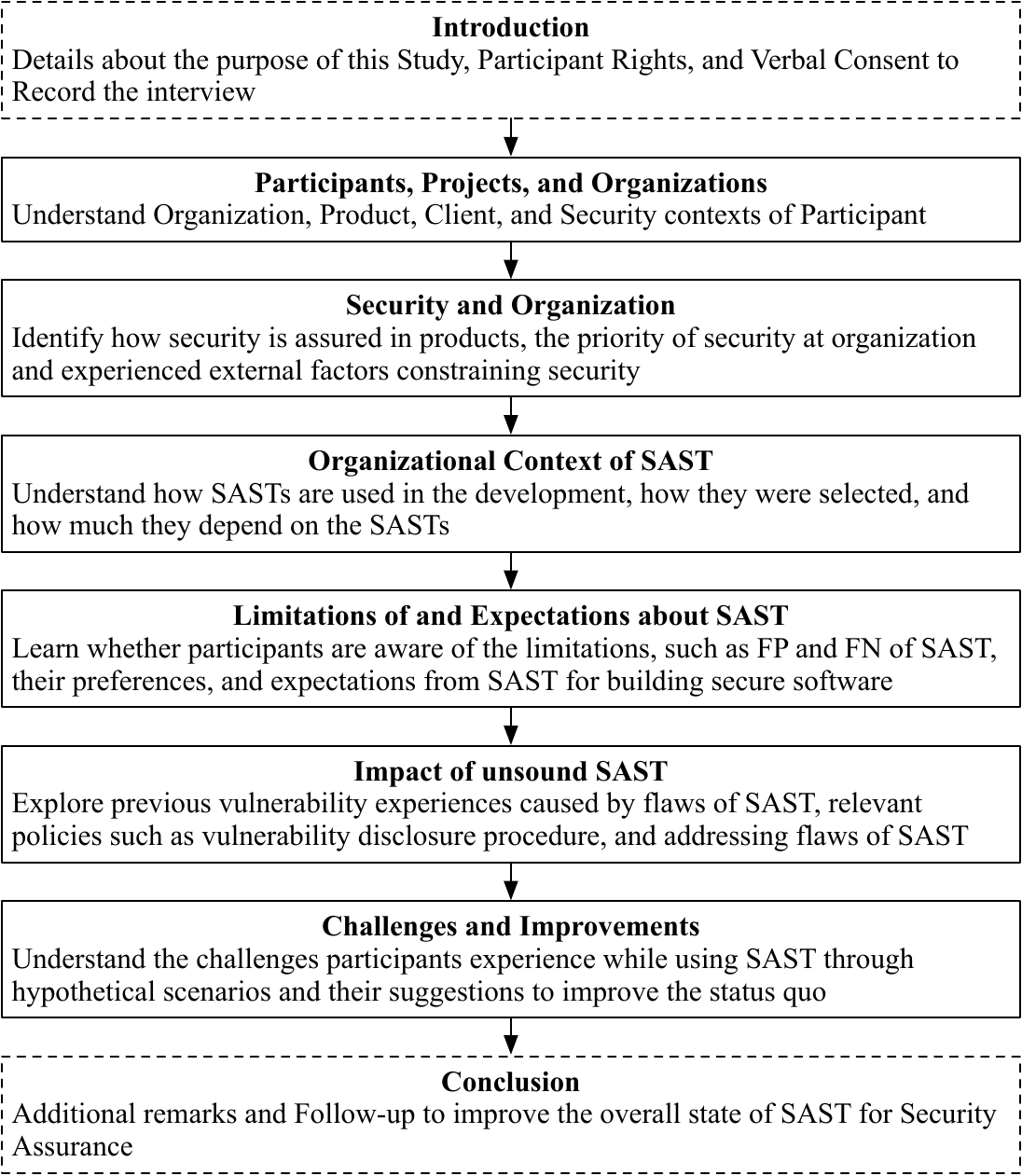}
	\vspace{-0.5em}
    \caption{Overview of Semi-structured Interview Guide. We used Laddering technique to delve deeper in each topic, while the semi-structured approach helped us to freely deviate as necessary based on participant response.}
    \label{fig:interview-guide}
	\vspace{-1em}
\end{figure}

We designed the semi-structured interview guide in a way that facilitates understanding how \add{practitioners at} organizations choose and depend on \sasts with the context of their business and security needs~(\rqref{1}), what practitioners with different needs and priorities know and assume about limitations, such as soundness issues, in \sasts~(\rqref{2}), how they address the limitations of \sasts{}~(\rqref{3}).
As shown in Figure~\ref{fig:interview-guide}, the interview guide consists of questions arranged in six segments, ordered by increasing-depth as applicable.

\subsubsection{\interviewSectionParticipants} At the start of the interview, we asked several "\textit{warm-up questions}" to understand the products/services the participants contribute to, their organization, their experience with developer tools for software security and how they define security in terms of their work.
Through these questions we developed the initial context to ask more in-depth follow-up questions.
More specifically, we asked the participants about their domain of work, their target clients, how they learned about software security that's relevant to their work, the security aspects that are important in their work, as well as the relevant threat models they consider. %

\vspace{-0.25em}
\subsubsection{\interviewSectionSecAndOrg} Next, to gain a deeper understanding of the organizational context of security in their practice, we asked the participants about how they address security in their organizations product development life-cycle. For example, we asked whether there are conflicts between feature deadlines and the security of a given feature, what the conflict resolution process is in general, and whether they have experienced any external factors that constrained security.
By raising such questions, we developed a better understanding of the trade-offs an organization makes when it comes to security. %

\vspace{-0.25em}
\subsubsection{\interviewSectionOrgOfSast} %
We then asked questions about how one or more \sasts are being used in organizational and team contexts.
From the survey, we observed that organizations and their developers may have different priorities and perceptions when it comes to security,  which motivated us to distinguish between these two contexts.
To elaborate, we asked the participants about their team structure, whether the team(s) address security requirements collaboratively or separately and how, and what happens when such requirements are not met.
Moreover, to understand the role of \sasts in the organization and team, we asked questions to understand how they decided to use \sasts in the first place, how they selected \sasts, and to explain why and to what degree they rely on \sasts.

\vspace{-0.25em}
\subsubsection{\interviewSectionLimitExpectSast}
We asked questions related to participant's expectations from \sasts and their limitations, using both
 hypothetical scenarios and also leveraging opinions the participant expressed throughout the interview.
For example, we asked about preferences regarding false positives vs. false negatives, explaining the concepts as necessary. Furthermore we asked whether their preference of SASTs is tied to their work, values shared within the community, or something else.

\vspace{-0.25em}
\subsubsection{\interviewSectionImpactUnsoundSast} To understand the impact of a flawed/unsound \sast, we asked the participants about their experiences and organizational processes.
For example, we requested that participants share specific experiences related to security vulnerabilities resulting from a \sast that did not work as intended.
If that participant did not have such experiences, we asked whether a process exists that helps them address such potential flaws.
In addition, we asked how and why the participants generally attempt to address problems encountered while using SASTs.

\subsubsection{\interviewSectionChallengeSolution} Finally, we concluded the interview by raising several "\textit{creative}" questions.
For example, if a participant is given unlimited resources to solve one particular problem of the \sast they use, what problem would they prioritize before anything else, and why would they want to solve it.
By raising such open-ended questions that \textit{remove limitations} tied to organization and product-context, we aimed to understand what participants want or need in the \sasts they use.

\vspace{-0.25em}
\subsection{Transcribing, Coding and Analysis}
\label{subsec:thematic}
\vspace{-0.25em}
One of the authors systematically transcribed the audio records while anonymizing the text.
This required significant amount of time as the median, effective interview duration was one hour, consisting of approximately $9,000$ words, with a total word count of over $187,000$ words across all interviews.
We chose reflexive thematic analysis combined with inductive coding for our analytical approach~\cite{braun2021thematic} as it offered us the flexibility of capturing both latent and semantic meaning based on the complex interactions between the participant's perceptions and contexts, such as assurances offered by automated security analysis techniques, organizational priorities, limitations of security resources, and the nature of products. Furthermore, it considers researcher subjectivity, \ie experiences and skills of researchers in analysis. We chose a single-coder approach, which is considered "good practice for reflexive TA", as it helps interpretation, or \textit{"meaning-making"}, from data~\cite{braun2021thematic}.
While transcribing provides an initial idea about the data and internal patterns, we had to iterate through the steps of thematic analysis (familiarization, coding, identifying potential themes, refining) to finalize the themes.

\begin{table*}[htbp]
	\centering
	\scriptsize
	\vspace{-1em}
	\caption{Overview of interviewed participants, position(s), product area(s), security priority from the perspectives of participants and project metadata.
		Fine-level details are binned to ensure the anonymity of the participants.}
	\vspace{-0.5em}
	\label{tbl:interview-participants}
	\def\arraystretch{1.2}
	\begin{tabularx}{\textwidth}{
			l   p{0.05\textwidth}  p{0.04\textwidth}  p{0.28\textwidth}      p{0.33\textwidth}      c c c}
		\toprule
		\textbf{ID} & \textbf{Duration}\tfootnotesymbol{1} & \textbf{\update{Channel}}\tfootnotesymbol{2} & \textbf{Position(s)\tfootnotesymbol{3}}     & \textbf{\update{Product Area}\tfootnotesymbol{4}} & \multicolumn{3}{c}{\textbf{Security Priority}}                                                                      \\
		\cmidrule(l{.5em} r{.7em}){6-8}
		            &                                      &                                              &                                             &                                                   & \textbf{Org}\tfootnotesymbol{5}                & \textbf{Dev}\tfootnotesymbol{6} & \textbf{Prod}\tfootnotesymbol{7} \\
		\midrule
		\multicolumn{8}{c}{Interview Participants Recruited from Survey}                                                                                                                                                                                                                                                          \\
		\midrule
		P01         & 01:00:12                             & OSS                                          & Senior Engineer                             & Program Analysis for Security                     & \ye                                            & \ye                             & \ye                              \\
		P02         & 01:04:08                             & OSS                                          & Developer                                   & OSS - Java Application Server                     & \ye                                            & \ye                             & \ye                              \\
		P03         & 00:51:04                             & OSS                                          & Developer                                   & OSS - Internet Anonymity Network                  & \ye                                            & \ye                             & \ye                              \\
		P04         & 01:00:44                             & OSS                                          & Embedded Engineer                           & Automobile Sensors                                & \ye                                            & \ye                             & \ye                              \\
		P05         & 00:57:46                             & PN                                           & Developer                                   & Web Applications                                  & \ye                                            & \ye                             & \ye                              \\
		P06         & 00:53:53                             & PN                                           & Developer                                   & Software Service                                  & \no                                            & \no                             & \ye                              \\
		P07         & 01:04:23                             & PN                                           & Engineering Manager                         & B2B, SaaS                                         & \ye                                            & \ye                             & \no                              \\
		P08         & 00:55:03                             & PN                                           & Full stack Developer                        & Media, Web and Back-end services                     & \ye                                            & \ye                             & \ye                              \\
		P09         & 01:05:25                             & PN                                           & Senior Engineer                             & Fintech, Business Critical                        & \ye                                            & \ye                             & \ye                              \\
		P10         & 01:01:40                             & PN                                           & Developer                                   & Healthcare                                          & \ye                                            & \ye                             & \ye                              \\
		\midrule
		\multicolumn{8}{c}{Interview Participants Recruited through Snowball Sampling from Professional Network}                                                                                                                                                                                                                  \\
		\midrule
		P11         & 01:01:48                             & SS                                           & Developer                                   & Website Backend of Program Analysis for Security  & \ye                                            & \ye                             & \ye                              \\
		P12         & 01:13:48                             & SS                                           & Developer                                   & Finance of International Online Marketplace       & \ye                                            & \ye                             & \ye                              \\
		P13         & 00:51:14                             & SS                                           & Research Engineer                           & Research Institute with Industry ties (EU)        & \ye                                            & \no                             & \no                              \\
		P14         & 01:02:09                             & SS                                           & Principal Configuration \& Dev-Ops Engineer & Law Enforcement                                   & \ye                                            & \ye                             & \ye                              \\
		P15         & 01:00:57                             & SS                                           & Senior Developer                            & Service Company                                   & \ye                                            & \ye                             & \no                              \\
		P16         & 01:03:43                             & SS                                           & AI Developer, Project Manager               & AI Products                                       & \ye                                            & \no                             & \no                              \\
		P17         & 00:49:15                             & SS                                           & Entrepreneur                                & Enterprise Resource Planning, Education platform  & \no                                            & \no                             & \no                              \\
		P18         & 00:50:53                             & SS                                           & Software Infrastructure Engineer            & Fortune 500 Global R\&D Center                    & \ye                                            & \no                             & \ye                              \\
		P19         & 01:05:13                             & SS                                           & Senior Software Engineer                    & Software Solution Provider                        & \ye                                            & \ye                             & \no                              \\
		P20         & 00:36:14                             & SS                                           & Backend Senior Software Engineer            & Telematics                                  & \ye                                            & \ye                             & \ye                              \\
		\bottomrule
		\multicolumn{8}{p{.98\textwidth}}{
			\tfootnotecontent{1}{Effective duration, \ie timed after introduction, briefing and verbal consent for starting to record the interview},
			\tfootnotecontent{2}{\update{Recruitment channel; OSS = Open Source Software developers, PN = Survey participants recruited from Professional Network, SS = Snowball Sampling within Professional Network}},
			\tfootnotecontent{3}{Self-reported by the participants, multiple roles separated by commas},
			\tfootnotecontent{4}{Product area binned, \update{none of the interviewees work in small (e.g., 2-person) organizations}},\
			\tfootnotecontent{5}{Invests in security in terms of the security team, tools, infrastructure, and/or training for developers (\ye)},\
			\tfootnotecontent{6}{Participant explicitly expressed that developers in organization are concerned about programmatic security, and/or is directly related with security tool development/setup (\ye) },\
			\tfootnotecontent{7}{Product is required to be compliant to privacy and/or security standards (\eg HIPAA, GDPR, PCI DSS, OWASP) or is critical in terms of safety (\ye).}
		}
	\end{tabularx}
	\vspace{-1.5em}
\end{table*}

\section{Interview Results}\label{sec:results}
This section describes the results from our interpretation and analysis of \countInterviews{} semi-structured interviews of practitioners, substantiated by transcribed quotes (omissions highlighted by \dots and anonymization in \angled{angle brackets}).
Also, when quoting participants (except when inline), we also include the product area, to provide further context behind the quote, \eg P20~\textsubscript{Telematics}.

\vspace{-0.5em}
\subsection{\interviewSectionParticipants}\label{sec:interviewSectionParticipants}
\vspace{-0.5em}
We succeeded in recruiting participants involved with a diverse range of products or services, such as web applications, anonymity networks, research software, safety-critical embedded systems, business-critical financial tech systems, and more, as illustrated in Table~\ref{tbl:interview-participants}.

All the participants in our study work on multiple projects, with a number of them working in multiple organizations.
Further, all the participants work in team(s), although the structure varies.
For instance, P01 worked in a team that maintains an Interactive Application Security Testing (IAST) product, but also used \sasts in their work.
On the other hand, P03 worked with a collection of people responsible for maintaining a popular open source anonymity protocol (\myquoteinline{It's a community run project...no corporation in control of any particular aspect of it}), and is responsible for configuring the \sast tools used by their team.

In summary, and as seen in Table~\ref{tbl:interview-participants}, the recruited participants possessed valuable, diverse experience by working in different types of projects with varying levels of security needs at their own, unique organizations.
This positioned us to further understand how productivity, in terms of feature implementation/completion, and security, in terms of ensuring that features implementations are not vulnerable, are balanced at these diverse organizations.

\vspace{-0.5em}
\subsection{\interviewSectionSecAndOrg}\label{sec:interviewSectionSecAndOrg}
\vspace{-0.5em}

Unsurprisingly, all participants agreed that delivering secure software is important.
However, we were also interested to learn about the prioritization of security in organizations.
Therefore, we queried participants about the potential tension within their organizations related to prioritizing software security at the expense of features and vice versa.
Given that introducing and using \sasts in a development workflow requires nontrivial effort from individuals and potential financial investment from an organization, we expected most participants and organizations had a vested interest in prioritizing security.
We found that this expectation to generally hold true, with a few exceptions.

\myparagraphnew{Prioritizing Security vs Functionality Deadlines}
Most participants indicated a prioritization of security over deadlines, \eg as P08 states, \myquoteinline{Security gets the highest priority. Always.\ldots Even if we are not meeting the deadline, we cannot break this.}
We found that various factors can be responsible for necessitating this prioritization, \eg the need to be compliant with existing laws and standards:
\myquote{We serve the government \dots we need to have some certifications that we are complying \dots (If) we have a release tomorrow and the security team found a vulnerability today, we have to block that release, and we have to fix it. Then we will release that. \dots We cannot compromise that.}{P20~\textsubscript{Telematics}}
It can also be due to safety-critical and/or business-critical nature of the product being built, since a bug can be costly, both in terms of lives and financial measures:

\myquote{So our security and safety and usage of the static analysis tools is mostly to prevent bugs, which could be life-threatening, of course, but also they could cost us millions}{P04\textsubscript{Automobile Sensors}}
For open-source collaborations, the concept of deadlines may not be applicable at all.
As P02 described, security is always of priority, and there is \myquoteinline{No such thing as deadlines. It's ready when it's ready}.

\finding{Participants generally said that they err on the side of security, fixing any known vulnerabilities before releasing a feature, regardless of deadlines.}

However, some participants expressed that prioritizing security is not always possible, even when security is generally a high priority from the organization's perspective.
This can be due to the management prioritizing bug-fixing for the sake of users, as shared by P07,
\myquote{\dots Our user was facing a lot of issues. So, there was a deadline pressure on us to deliver the product very quickly}{P07~\textsubscript{B2B, SAAS}}
This is true even for a security-testing product, albeit rarely:
\myquote{So in most cases, we try to be really strict because it's a security testing product \dots Then it's kind of a business trade off. \dots a new feature, we can usually just delay it \dots If it's an existing feature that we now uncover the vulnerability, you can't usually switch off the feature because you have customers relying on it.\dots Eventually you get that fixed and then responsibly disclose it. \dots I think we had to do one of those in the five years I've been with the company.}{P01\textsubscript{Program Analysis for Security}}
The "overriding" of security to meet deadlines for existing features may incur heavy cost, however. P07 expressed the following after further conversation, \myquoteinline{That had a certain impact. We found \underline{.3 to .4 million <currency>} of fraudulent activity after that release}.

\finding{Select situations can lead to the de-prioritization of software security, including maintaining support for {\em existing} features, or fixing bugs that being experienced by prominent users.}

\add{This finding echoes similar observations in prior work, that in some cases security is forgone for functionality bugs or releasing other features~\cite{FCV+21,PTL+20,AC19,XWM14}.}

Further, contrary to our initial intuition, P06 shared that an organization may not prioritize security unless it is required by its clients. %
\myquote{Security is a great concern \dots So if the client is strict enough to focus on the security aspects, then we follow it. Other than that, actually our <previous org> do not care (about security) \ldots}{P06\textsubscript{Software Service}}
Unsurprisingly, several participants shared that an organization may not afford to miss functionality deadlines if it is still in startup or growth stage:
\myquote{When I worked on a startup environment, it is always expected that we ship the features to production as soon as possible...There is a little room to explore the security options \dots}{P10~\textsubscript{Healthcare}}
\finding{Participants expressed that in certain circumstances, organizations may entirely forego security considerations and prioritize releasing features as rapidly as possible, particularly when the client does not care, or when the organization is in its early growth stage.}

\subsection{\interviewSectionOrgOfSast}\label{sec:interviewSectionOrgOfSast}
\vspace{-0.5em}
With the understanding of how security is viewed from a participant's perspective in the context of their organization, we aimed to understand their organizational context of \sasts.
Particularly, we asked participants about reasons for using \sasts{}, the selection process of \sasts, and how they generally use those \sasts in their workflows.
Although participants named \sasts, \eg Coverity, Find-Sec-Bugs, SonarQube, Semmle/CodeQL, WhiteSource/Mend, CryptoGuard, Fortify, and VeraCode, we anonymize such details to reduce the chance of profiling developers and to avoid creating any impressions specific to any particular \sast.

\myparagraphnew{Selecting \sasts}
To start, we asked participants about the events that led to choosing a given \sast{} and to walk us through the selection process at their organizations.
Interestingly, we did not find any pattern in the selection processes that would hold true for a majority of the participants.

Particularly, only P04 shared that they performed a multi-stage evaluation, \ie they started with a preliminary list of $10-15$ \sasts, and filtered to four \sasts based on their own product-specific needs. Next, they evaluated those four \sasts using a {\em custom} benchmark, and settled on a \sast that was the most usable.

Further, six practitioners shared that they chose \sasts solely based on popularity, developer friendly documentation and/or ease of use.
\myquote{We didn't evaluate that many tools in terms of static analysis tools. We take what is the industry standard across different companies. Like <tool> is pretty popular, so that is our first choice.}{P08\textsubscript{Media, Web and Back-end servicess}}
P09 additionally mentioned that \textit{<SAST\textsubscript{A}>} was chosen due to regulatory reasons, \myquoteinline{I believe it was either PCI DSS requirement or a regulatory requirement}, admitting that they do not remember the exact standard.
On the other hand, four participants reported using previous experience or familiarity to select a \sast.
\myquote{Because I actually inherited some of that. The person who, actually set a lot of it up ... I think that it was what was available and what he was familiar with at the time.}{P03\textsubscript{OSS - Internet Anonymity Network}}
Several developers justified that they prefer freely available \sasts because it helps cut costs,
\eg P07 said \myquoteinline{...As of now, we are looking at a free solution. If we find benefits, then we'll go for the paid solution ...}

Corporate influence is an additional factor for selecting a particular \sast, particularly when it comes down to cost, \eg as P14 said, \myquoteinline{A lot of it comes down from management, \dots because they're the ones that are paying for it}.
In a similar vein,
P08 shared:
\myquote{We have different teams and different teams have different requirements. In my team, we use <SAST\textsubscript{A}>, and it is enforced by the team leader or the team owner to use <SAST\textsubscript{A}> as a code analysis tool.}{P08\textsubscript{Media, Web and Back-end Services}}
\finding{Participants generally recall selecting SASTs due to factors such as recommendations/reputation, ease of use/integration, corporate pressure, cost, or compliance requirements. Only {\em one} participant selected a SAST for their product via exhaustive testing of 10-15 tools using a (custom) benchmark.}
Furthermore, we asked participants about whether they considered using benchmarks, such as the OWASP benchmark while selecting \sast.
Most participants said that they were not familiar with any benchmarks, with the rest sharing that benchmarks are not representative of their specific application context, \eg \myquoteinline{The thing is, OWASP is something that only covers your basics. It doesn't go beyond} (P09).
Furthermore, P01 shared that while community-based benchmarks such as OWASP are usually neutral, many others are biased.
\myquote{%
Quite a few of these benchmarks are created by tool vendors where their tool finds some specific edge case. No one in the right mind would write an application like this, but their tool finds a specific edge case, so they put it in the benchmark.}{P01\textsubscript{Program Analysis for Security}}
\vspace{0.25em}
\finding{Participants who are aware of benchmarks generally do not trust them for evaluating/selecting SASTs, viewing benchmarks as either too basic to model real problems, or biased towards specific SASTs, given that vendors often contribute to their construction.}
\myparagraphnew{Preference between Manual Techniques and \sast{}}
As expected, participants who use \sasts stated that they found them useful, regardless of the selection process. %
Several participants shared that they use \sasts because they help focus manual analysis efforts on non-trivial issues by \textit{automatically} finding the trivial issues, \eg \myquoteinline{\dots helps find all the stupid stuff for you. Then you can concentrate on the actual logic (P01)} and makes it easier to analyze a large code base, \eg \myquoteinline{Is it possible to go through each of the code change by a human being? (P20)} and \myquoteinline{I think they're absolutely useful. It kind of reduces the number of mistakes you can make} (P09).
Furthermore, several shared that it is helpful for applying a rigorous quality control to the whole code base without being affected by subjective analysis, \eg
\myquote{Lot of reasons to be paranoid about it. None of us really, totally trust ourselves. And so, we need to have these tools to make the job of finding our own mistakes easier. If only one person is working on a thing, you're stuck with only that person's blind spots.
}{P03\textsubscript{OSS - Internet Anonymity Network}}
\finding{Participants consider SASTs highly useful for both reducing developer effort and helping to cover what subjective manual analysis may miss.}
\myparagraphnew{Reasons for not relying on \sast{}}
Finally, we had two participants in our study who do not rely on \sasts.
P13 stated that while their product needs to be secure, it is not public-facing, \ie \myquoteinline{Even if there is a problem in some projects, so one can access those deployed or the application from outside of our internet}.
Interestingly, P02 shared that while they have tried premium \sasts, they did not find them useful in their particular application niche, \ie web servers, stating that
\myquote{The primary issue with the <generic SAST> tools, every time we've looked at these tools, is it's all false positives and no genuine issues at all, which is somewhat demoralizing if you try to wade through large amounts of these reports.
}{P02\textsubscript{OSS - Java App Server}}
That is, as P02 further elaborated, since their product is a web server, it is required to handle "vulnerable" requests, such as "HTTP" headers, in code based on existing standards. These code components, however, trigger \sasts built to target web-applications, resulting in high false positives.

\finding{\update{The few participants who do not use \sasts{} cite the lack of a ``{\em fit}'' for their product: \ie{} as the product does not need extensive testing (echoing similar observations in prior work~\cite{WZW+15}), or because generic \sasts{} flag features (\eg handling standard-mandated vulnerable HTTP requests) as vulnerabilities.}
}

\vspace{-0.5em}
\subsection{\interviewSectionLimitExpectSast}\label{sec:interviewSectionLimitExpectSast}
\vspace{-0.5em}
Developers shared that while they expect \sast tools to detect all vulnerabilities as long as they are within scope, they generally do not expect \sasts to detect all types of security vulnerabilities, \eg \myquoteinline{If it is in the scope, then it can detect, but my expectation is not like static analysis is the final solution...(P10)}.
When asked whether this assertion was based on ``belief'' or ``evidence'', P18 explained that it was \myquoteinline{based on belief}, further explaining that \myquoteinline{We have the user ratings of our tools and there are many stars in the repos. So we think that it is reliable, and many developers use that, so it must be good}.

\finding{Although expressing that no tool can find everything, participants {\em believe} that SASTs (should or do) detect {\em all vulnerabilities considered \underline{within scope}} (\ie which a SAST tool claims to detect).}
When asked to give examples of vulnerabilities that developers do not expect \sasts to detect, runtime (Input/Output), external component, and software goal based issues frequently came up, \eg \myquoteinline{These tools are pretty agnostic of the goals that we have put forward in the first place. They can only really seem to process errors in code and not errors in software taken as a whole (P03)}.

Interestingly, when we asked developers if they consider a \sast{} to be acceptable to use even if it misses some more difficult issues, they generally expressed that they do, \eg \myquoteinline{I don't expect that there will ever be a tool that will
look at a piece of code as complex as \angled{product} and find all the security issues. \dots But any issue fixed is an issue fixed and that's a good thing} (P02).
When asked to elaborate, participants shared different reasons for finding such \sasts acceptable, such as lack of alternatives, \myquoteinline{If there is no other accessible alternative, then I would go and accept whatever it offers} (P10) and additional techniques being used to cover for (issues in) \sasts \eg \myquoteinline{\dots for our team, the manual review part is actually the biggest deal for us. \dots So, for our team, I think that should not be a big issue} (P07).

\finding{Participants consider SASTs valuable even if they miss certain vulnerabilities, as {\em finding something would be better than nothing.}}

\myparagraphnew{Reducing False Positives vs False Negatives}
In the context of program analysis, increasing analysis sensitivity decreases false negatives, while increasing false positives, and vice versa.
Contemporary literature asserts that false positives are a major reason for practitioners to avoid using SASTs~\cite{JSMB13, IRFW19} since \textit{"Developer Happiness is Key"}~\cite{SAE+18}, and argues that it is necessary to reduce false positives, \textit{in general}.
That is, conventional wisdom dictates that developers want lower false positives even at the cost of false negatives, which has led to a significant focus on increasing the precision of SASTs in academia and industry in pursuit of practicality~\cite{RXA+19, 2018_AmandroidPreciseGeneral_wei, 2013_FlowDroidPreciseContext_arzt, FAR+13, KFSZ18, 2015_ScalablePreciseTaint_huang, AB16}.

However, we found considerable evidence that contradicts this understanding of the developers' perspective on the soundness-precision tradeoff, with participants strongly favoring lower false negatives, even at the cost of increased false positives. As P04 and P06 state,

\myquote{False negative for sure. I just told you the amount of the price of the bug (in millions), so I don't care if there are 10 false positives. False negative - that one is going to kill you.}{P04\textsubscript{Automobile Sensors}}
\myquote{From my understanding it is actually more threatening that we aren't even aware of the vulnerability...So to me, false negative should be bigger concern...it (false positives) wastes time of developers, but it is not harmful in the whole picture}{P06\textsubscript{Software Service}}
P14 even argued that false positives indicate a working \sast, and when it comes to security, no stone should be left unturned, \myquote{If you're getting a bunch of false positives, then that typically means your static code analysis tool is doing its job. \dots I'd rather my security tool be annoying and tell me about every single possible issue over it not telling me anything and just letting security things slide through.}{P14\textsubscript{Law Enforcement }}
\finding{Nearly all the practitioners expressed a preference for {\em fewer false negatives}, \ie as long as the \sast is able to find valid security vulnerabilities, they would tolerate and even prefer few false negatives at the cost of many false positives.}
Since existing literature argues that false positive rate for program analysis should not exceed 20\%~\cite{JSMB13, SFZ11, CB16, BBC+10 }, we requested our participants to approximately quantify their preference (or experience) regarding the acceptable proportion of false positives to true positives.
For most participants, this preference was far higher than 20\% as long as the tool detected some valid vulnerabilities (i.e., had true positives), as indicated in \fnumber{10} as well.
For instance, P02 admitted to dropping a tool in favor of manual analysis due to overwhelming false positives without a single valid vulnerability:
\myquote{I wouldn't mind wading through $100$ false positives, if I thought there were actually going to be genuine issues there}{P02\textsubscript{OSS - Java App Server}}
Some participants expressed tolerance for 80\% or more false positives, although not to the extreme extent as P02.
\myquote{The acceptable range is for (reducing) one false negative, that there could be five false positive}{P10\textsubscript{Healthcare}}
Further, some, \eg P09, stated that 80\% false positives were common in a tool they were currently using; although they were dismayed by the low number of serious, real vulnerabilities found:
\myquote{(At present) 80\% of them are actually false positives and 20\% of them are actually something we can fix. Even those 20, you don't generally find serious problems.}{P09\textsubscript{Fintech}}
Finally, P01 expressed a lower tolerance for FPs than most other practitioners, stating that \myquote{We ended up with $20\%$ real issues. $80\%$ just false positives. And one of my last actions in that company before leaving was saying, `Hey, look, this tool is a waste of time'}{P01\textsubscript{Program Analysis for Security}}
\finding{Practitioners are generally more tolerant of false positives than the 20\% upper bound proposed in literature, given their preferences and the tools they currently use, with some finding even 80\% or more false positives practical.}

\myparagraphnew{Effective False Positives and \sast}
Due to the perceived notoriety of false positives affecting adoption of general static analysis tools, the notion of \textit{effective false positives}, defined as \textit{"any report from the tool where a user chooses not
to take action to resolve the report"}~\cite{SVJ+15a}, or in other words - letting the developer determine whether {\em any} reported defect is a false positive, is gaining attention.
Effective false positives have further been contextualized in \sasts in the form of letting a developer determine whether a reported vulnerability should be considered as within the scope of security context~\cite{WBS+22}.
However, several participants cautioned that in their experience, developers may not make the right call when it comes to identifying an effective false positive issue \eg when an insecure code segment is considered "\textit{inactive}".

P14 and P15 expressed something similar to "\textit{The Developer is the enemy}" threat model~\cite{Wv08}. P15 argues that \myquoteinline{Junior developers don't understand what is the impact}, and P14 states (on effective false positives):
\myquoteinline{From a security standpoint, you can't really trust, you shouldn't trust other devs, and users to always know that something could be potentially insecure. So you need to make sure that it's not possible for it to happen or reduce the possibility of it happening as much as possible (by removing insecure code)}.
Furthermore, P14 cautioned that developers may habitually mark an actual issue as false positives erroneously, \myquote{It seems familiar, but it may be new. And then you're just going to ignore it because it's close enough to something you've seen in the past, and you just say that it's OK. So we do need to be vigilant on those false positives to make sure that they are truly false positives}{P14\textsubscript{Law Enforcement}}
P04 made a similar remark about making mistakes in deciding whether to run \sast or not on code patches, sharing that they had a vulnerability that could've been detected using static analysis, but was not due to the deliberate decision of  not using \sast, costing millions:
\myquote{The undisclosed amount is in a couple of millions. \dots  We had two static analysis tools which should be used, but the decision from the management was because it was a minor fix that they did not use them}{P04\textsubscript{Automobile Sensors}}
\finding{Practitioners are generally against letting developers define ``effective'' false positives, or letting them decide when to run SASTs. This reservation stems from their prior experience of the adverse cost of leaving a vulnerability in the code, and/or from their knowledge of developers (1) lacking an understanding of the impact of vulnerabilities, (2) being prone to incorrectly marking actual issues as false positives, (3) being untrustworthy/biased towards marking issues as effective false positives.}

\vspace{-1.5em}
\subsection{\interviewSectionImpactUnsoundSast}\label{sec:interviewSectionImpactUnsoundSast}
\vspace{-1em}
After learning about what participants expect from SASTs, we aimed to understand if and how participants were impacted by flaws in \sasts, \ie their inability to detect what they claim as ``in scope'', how participants generally addressed the flaws, and their experiences reporting the flaws to SASTs.

\myparagraphnew{Impact of Unsound \sast{}} All developers across survey and interviews, save for a few, shared that while they had experienced  false negatives, they had not experienced any adverse impact due to flaws/unsoundness in \sasts.
The practitioners explained that while false negatives are not observable since they are not reported by the SAST, they expect manual/code reviews to detect vulnerabilities missed by SASTs.
Therefore, as any false negatives resulting from even unknown flaws in SASTs are addressed by their manual reviews, unsound \sasts do not impact their software.

\finding{Practitioners are not overly concerned about the impact of unknown unsoundness issues in SASTs, as they expect subsequent manual reviews to find what the SAST missed.}

P18, who works with an internal static analysis team, offered an alternate explanation as to why developers may overlook false negatives of SASTs, or their impact, because the assumption is that \sasts \textit{just work}
\myquote{If the tools miss something, we can not detect that issue, and we just overlook the issues\dots because no one ever reports about false negatives, and we don't check if the tool ever miss the vulnerabilities}{P18~\textsubscript{Fortune 500 Global R\&D Center}}
\finding{Developers may use SASTs in a state of denial, i.e., assume that SASTs just work, and hence, simply overlook any evidence of false negatives, or flaws in the SASTs that lead to false negatives.}
Among the exceptions, P02's organization tried and stopped using \sast because of false negatives, thus effectively negating any potential impact, as previously described in~\ref{sec:interviewSectionLimitExpectSast}.
On the other hand, P01 shared that while their own \sast product unintentionally introduced a vulnerability, which could've impacted their clients, \myquoteinline{never public, no customer ever suffered}, as it was detected during development.

\myparagraphnew{Addressing/Reporting flaws to \sasts} Participants expressed that security is important, but shared challenges associated with reporting flaws to \sasts.

Generally, flaw reports consist of either a minimal code example that demonstrates the flaw, or actual code snippet from software.
However, P04 and P09 shared that going for either is problematic for two very different reasons. First, sharing actual code snippet may require going against company or client's  confidentiality policy. P04 circumvents this because of a pre-existing NDA between their organization and the \sast, \myquoteinline{we have an NDA signed, so in case I cannot get a small example, they can also check our source code}, whereas P09 is unable to do so.
\myquote{For certain external communications, it's a little bit difficult to do. What we can share with third party or other party is very strictly regulated by the state bank\dots. If we want something from <tool>, we have to justify why we are sharing this particular code snippet. In particular, I think if you don't share a large amount of code with them, they won't even be able to tell why this is problematic}{P09\textsubscript{Fintech}}
On the other hand, several participants stated that sometimes, developers are not willing to report flaws since it is "\textit{additional work}" (\ie reporting the flaw, following up): %
\myquote{We were asked to not do things on our own, because
they will maybe increase more pressure \dots  I would actually report it to my team lead, but I
don't think they would actually report it to back to them}{P05\textsubscript{Web Applications}}
\myquote{That might not happen as well because inherently developers are lazy. If you want to share this, you have to go through with certain things}{P11\textsubscript{Website Backend of Program Analysis for Security}}
Finally, some participants shared that while they have reported flaws to \sasts, the lack of response, or lack of addressing flaws discouraged them from reporting flaws later on.
P02 said, \myquoteinline{Nothing as far as I recall} when asked about whether anything happened after reporting false negatives to \sast, whereas P04 said that some \sast developers might be unwilling to accept a flaw as an issue.
\myquote{So, <SAST\textsubscript{A}>, we have a worse experience. They are mostly evasive, so they are not really progressing as <SAST\textsubscript{B}>. It takes a lot of time to convince them that they are bugs. Even though you have a small example, they still ask you to try different configurations and all that stuff, but we were aware of that before we came to this part, before we selected them. Because simply they (<SAST\textsubscript{A}>) are, I wouldn't say confident, but they are confident that their solution works.}{P04\textsubscript{Automobile Sensors}}

\finding{Participants may hesitate to report flaws/false negatives in SASTs for several reasons, ranging from prior negative experiences with SASTs (including inaction on reported flaws), or issues internal to the organization, such as the need to maintain product confidentiality (without an explicit NDA), red tape, and the lack of incentive to perform the additional effort.}

P01 shared some insight to decisions related to fixing flaws in \sast, sharing that while severity and likeliness ("correlates to presence in open source libraries") are motivating factors, so is what the business-competitors are detecting.
To understand this in-depth, we presented a hypothetical scenario to P01 where a class of vulnerability is ignored by the rest of the \sast building industry and asked how is it decided whether to address it in their \sast. P01's response was \myquoteinline{It depends on the effort and depends on how critical it is}.

\myparagraphnew{Exploiting Flaws and Evasive Developers} We adopted the concept of evasive developers from ~\cite{ACK+22, Wv08}, defined as a developer who actively attempts to bypass a \sast's checks.
The motives vary, such as malice, lack of stake (third-party contractor), and/or simply being lazy.
A majority of the participants stated that while they consider evasive developers realistic, such developers are unlikely to cause serious harm in their organizational context due to several factors, such as company policies \eg \myquoteinline{It is strictly prohibited, and it is communicated in that way that it is not acceptable to bypass those checks (P08)}, and manual code reviews.
\myquote{The process that we have is designed that, first, it needs to pass the review of the initial reviewer which allows it to get it on the main branch. So if we, put another hurdle here and we say that there are two friends which decide that this is okay, it still needs to come through the third guy who is gonna test, the test will kill. So that's already three guys that would need to accept the issue in the whole team.}{P04\textsubscript{Automobile Sensors}}
On the other hand, some participants shared that they have observed their colleagues being evasive, or they themselves attempted to be evasive due to stressed work environment.
\myquote{We had six people and one would actually do something like that.}{P05\textsubscript{Web Applications}}
\myquote{There was an extreme pressure because we \underline{needed to bypass the \sast tests}, otherwise we would not receive green flag from the security team. So it actually \underline{happened once}. We used to work late night to resolve all those conflicts and red flags.}{P06\textsubscript{Software Service}}
In contrast, P01 expressed that in an organization a developer being evasive is unlikely due to ownership at their organization, \myquoteinline{I want to believe that our developers are responsible \dots I don't believe anyone will try to game our system like that}.

\finding{The risk of evasive developers is real. That is, while some participants consider the scenario of ``evasive developers'' as adequately prevented by existing code reviews, this optimism is not universal: others have prior experience of evasive developers in their teams, or have evaded SASTs themselves.}

\subsection{\interviewSectionChallengeSolution}\label{sec:interviewSectionChallengeSolution}

Finally, we wanted to learn about the pain-points of participants related to \sasts.
Our approach was to present hypothetical (but ideal) scenarios, such as unlimited resources to fix or address just one issue of \sast, with the goal of getting the participants to focus on the most severe \sast-specific issues in their perspective. %

A few participants wanted to invest their resources on improving analysis techniques, both for reducing false negatives \eg \myquoteinline{I guess the first thing would be I would try to make it so that we're covering all of the most obvious} (P14), and providing meaningful alert messages \eg
\myquote{So the static analysis tool should be able to detect all the security issues within its scope and within possibilities. It should show meaningful messages \dots it should expose enough information about the issue so that the respective developer can address the issue easily}{P10~\textsubscript{Healthcare}}
Alternately, P02 (who was generally unimpressed by SASTs throughout the study) wanted unlimited human resources for manual analysis:
\myquote{If I've got unlimited time and resources, then some poor, unfortunate soul is going \dots going to have to go through all of the false positives in <SAST> and just confirm that they are actually false positives because there's just so many of them. \dots If those unlimited resources included some experienced security researchers, I get them doing some manual analysis. Because to be perfectly honest, the best vulnerability reports we get, which generally tends to be the more serious issues, they're not found by tools, they're found by people}{P02\textsubscript{OSS - Server}}
Several participants focused on \sast CI/CD integration issues explaining that often configuration is a major pain-point for them \eg \myquoteinline{I would definitely say integrations would be the top. \dots I think the best example would just be for all major CI/CDs to have an open source example of how to implement and integrate with various things.}. Other responses covered niches, such as better language-specific support, concurrency and abstraction support.

Finally, participants generally agreed that actionable reports that explain what can be done to address an issue, or provide more context, would be useful \eg \myquoteinline{if you write this code like this, this issue should be resolved} (P12), and \myquoteinline{An explanation of why the tool flagged that particular code is very helpful. It saves us having to second guess on why is the tool reporting that} (P02).

\finding{The key pain points for developers when it comes to SAST tools include: false negatives, lack of meaningful alert messages/reports, and configuration/integration into product CI/CD pipelines.}

        \vspace{-0.5em}
\section{Threats to Validity}\label{sec:limitations}
\vspace{-0.5em}

\add{
This study seeks to understand the diverse perspectives of practitioners with different types of business and security needs, and is affected by the following threats to validity:

\myparagraphnew{Internal Validity} Practitioners with different experiences and roles at organizations may provide responses influenced by over/under-reporting, self-censorship, recall, and sampling bias.
We mitigated these factors by asking participants to share organization-specific incidents and experiences, with follow-up questions to understand their context, and reassuring that the responses would remain anonymous and untraceable (Section~\ref{sec:interview-guide}).
Moreover, some participants may have experienced loss of agency in selecting \sasts{} (\eg P08, \fnumber{4}).
However, all our participants have played key roles in selecting {\em or} using \sasts in their organizations (see Section~\ref{sec:interviewSectionOrgOfSast}), leading to useful experiences and observations that reveal meaningful patterns in SAST selection.

\myparagraphnew{External Validity} Due to the nature of interview-based qualitative research focusing on a specific experience (here: with \sast{}), \textit{generalizability} is considered an issue for recruitment through snowball/convenience sampling.
Findings from such studies are considered \textit{"softly generalisable"}~\cite{braun2021thematic}.
Prior research demonstrates that such studies are reliable for identifying salient trends~\cite{AB96,DLB05}; indeed, given the diverse organizational and product contexts of our participants, their responses provide key insight into how \sasts{} are used in practice in complex organizations.

In other words, given the number of our participants (n=20), and the recruitment process, we do not claim that the participants are representative of the broader developer population, or that the findings are \textit{generalizable}.
That said, this study captures and analyzes the experiences of participants from diverse organizational and security contexts and offers salient insights related to the use of \sasts{} in practice.

}

    \vspace{-0.5em}
\section{Discussion}\label{sec:discussion}
\vspace{-0.5em}

The findings from our study reveal salient aspects of how developers use SASTs, what they expect from them, and how they react when SASTs do not fulfill those expectations.
We now distill the findings into four themes related to the problems inherent in the use and perceptions of SASTs as well as a path forward for researchers and practitioners.

\vspace{-0.5em}
\subsection{Mind the Gap: The Dichotomy of Perceived Developer Needs and SAST Selection/Evaluation}
\vspace{-0.5em}

A common sentiment observed throughout this study is that practitioners do care about and prioritize security.
To elaborate, practitioners stated that they would generally fix vulnerabilities regardless of release deadlines (\fnumber{1}), except in certain mitigating circumstances (\fnumber{2}, \fnumber{3}), and use SASTs to cover the blind spots and subjectivity pertinent to manual code analysis (\fnumber{6}).
Moreover, nearly all practitioners favored lower false negatives (\ie not "letting security things slide through") (\fnumber{10}), expressing a surprising tolerance for false positives as long as SASTs found vulnerabilities ~(\fnumber{11}).

However, we found that this strong preference for security, and particularly SASTs that find real vulnerabilities, is not reflected in how practitioners select SASTs.
To elaborate, practitioners select SASTs based on cost, corporate pressure, ease of integration/use, and particularly, recommendations from peers and general reputation of the tool (\fnumber{5}).
This generally ad-hoc and subjective criteria does not provide objective evidence of a SAST's performance in detecting vulnerabilities.
Hence, there is a clear {\em gap} between the criteria that practitioners use for selecting SASTs, and what they want most from SASTs (evidence of real vulnerability detection abilities).

\subsection{The Power of Reputation and the Lack of Reliable Objective Criteria}

The key question is, {\em why does  this gap exist? That is, why don't practitioners evaluate the security properties of SASTs?}
Our findings point to two key reasons:

First, we find that practitioners may not have any motivation to evaluate SASTs.
That is, practitioners seem to be unreasonably optimistic about the SASTs' abilities, assuming that SASTs must detect everything they claim to (\ie define as within scope) (\fnumber{8}), and assume that SASTs "just work" (\fnumber{14}).
This optimism, coupled with their reliance on reputation as a valid metric for selecting SASTs (\fnumber{4}), may be sufficient to dissuade practitioners from any additional effort required to evaluate SASTs.
Thus, the observed lack of motivation to evaluate SASTs is concerning, particularly as the blind belief practitioners express in SASTs and their reputation does not hold up to scrutiny: \eg a recent evaluation of reputed crypto-API vulnerability detectors showed serious, previously unknown flaws, which prevent the detectors from finding vulnerabilities they consider ``in scope''~\cite{ACK+22}.

Second, even when practitioners want to evaluate SASTs, the existing means to do so, \ie benchmarks, are perceived as insufficient.
As we found, while some practitioners may be unaware of benchmarks for evaluating SASTs, most are not.
In fact, most practitioners {\em do not trust existing benchmarks}, viewing them as either too basic (and not representative of real, complex, vulnerabilities), or biased (\fnumber{5}).
These findings indicate a significant gap in the research on evaluating SASTs, and motivate the development of high-quality, comprehensive, real-world benchmarks vetted by both researchers and practitioners, if we intend to help practitioners objectively evaluate SASTs for what they most desire: the ability to detect vulnerabilities.

\vspace{-0.5em}
\subsection{Giving Developers What They Want}
\vspace{-0.5em}
We observe that practitioners repeatedly expressed that they want two things from \sasts: ease of configuration (\fnumber{4}, \fnumber{17}), and for tools to detect real vulnerabilities (\fnumber{8}).

Fortunately, the ease of configuration is being addressed by the recent, additional support for \sasts through integration into CI/CD pipelines of open source projects, such as via Github Actions~\cite{github-actions, github-code-scan}, as well as standardized output formats, such as SARIF~\cite{SARIF}.
However, the latter is harder to achieve at present.
That is, while our practitioners repeatedly expressed that they want SASTs to be first and foremost able to find critical vulnerabilities (and all those considered within scope, \fnumber{8}), even at the cost of higher number of false positives (\fnumber{10}, \fnumber{11}), the research community continues to show preference towards improving precision instead, \ie decreasing false positives, for SAST tools throughout the last decade~\cite{2018_AmandroidPreciseGeneral_wei,2013_FlowDroidPreciseContext_arzt,2015_ScalablePreciseTaint_huang,RXA+19, BBC+10, SFZ11,SAE+18}. %
Thus, for SASTs to actually be useful, the research and industry communities  need to refocus their efforts towards finding critical vulnerabilities (and all that is deemed within scope), with improved precision being an additional, desired, property.

\vspace{-0.5em}
\subsection{Industry is not prepared for the flaws of \sasts{}}
\vspace{-0.5em}

Our findings expose a {\bf \em critical paradox} in the assumptions industry practitioners make about their approach towards SASTs:
While practitioners do expect SASTs to detect all vulnerabilities within scope (\fnumber{8}), they are not overly concerned with SASTs missing such vulnerabilities due to undocumented flaws, because their subsequent manual analysis to find what the SASTs missed (\fnumber{13}).
However, practitioners also emphasized that their key reason for using SASTs is to account for knowledge gaps, blind spots, and subjectivity inherent in manual analysis (\fnumber{6}, \fnumber{12}).
To summarize the paradox, practitioners use SASTs to account for gaps in manual analysis, but then, in turn, are confident that manual analysis will account for (unknown) \sast flaws.

This paradox suggests several undesirable aspects of the status quo. First, that developers may be overly confident in guarantees offered by their process of combining SASTs (or other tools) and manual analysis, or may simply take the reports of SASTs at face value (\fnumber{14}), which may result in undetected vulnerabilities in code that are missed by both SASTs and manual analysis; \eg previous work has shown that the same undocumented flaws can manifest in any number of \sasts, and lead to vulnerabilities in programs analyzed by the SASTs~\cite{ACK+22,AKM+21}.
Second, given that practitioners generally hesitate to report flaws in SASTs (\fnumber{15}), the flaws in a SAST would persist and harm most software using the SAST, even if a few practitioners do uncover false negatives/flaws during manual analysis.
That is, if the status quo observed in this study continues, SASTs will likely never improve in their ability to detect vulnerabilities, but instead, will continue to be used in a manner that inspires a false sense of security among practitioners.

To summarize, we conclude that the industry is ill-equipped to find or address any flaws in \sasts, particularly given the state of current reporting processes that are mired in confidentiality issues, an evasive attitude, and lack of response from \sasts (\fnumber{15}).
Thus, practitioners are stuck with repurposing common issue submission processes that does not cater to their confidentiality needs, does not elicit a response, and does not facilitate discussion.

\vspace{-0.25em}
\subsection{Moving Forward: New Directions and Ideas}
\vspace{-0.25em}

To improve this status quo, researchers and practitioners need to establish a dedicated process for reporting false negatives, as well as expectations from SASTs upon receiving such reports, in a manner similar to bug reporting expectations for typical software products. This might involve the development of automated methods for creating minimal examples of vulnerabilities missed by SASTs or even ``self-healing'' SASTs that leverage advancements in automated program repair to address missed vulnerabilities.
Moreover, future work may also explore streamlining the automated evaluation of SASTs (\eg developing web-based services that allow practitioners to "test" SASTs with realistic vulnerabilities), so that developers may be able to evaluate SASTs before using them, instead of leveraging subjective criteria for the same.
Beyond evaluation techniques, researchers should also consider orienting future work on SAST development toward the high preference of practitioners in finding important/critical vulnerabilities, even at the expense of a high number of false positives. 

To summarize, only by raising awareness about the flaws in \sasts, aligning their goals with the goals of developers, designing protocols for evaluating them, and streamlining bug reporting, particularly for false negatives, can we move towards a more desired state where practitioners are able to leverage SASTs to their true potential, resulting in a holistic reduction in hard-to-find vulnerabilities.

    \vspace{-0.25em}
\section{Related Work}\label{sec:relatedwork}
\vspace{-0.25em}

SASTs have been adapted for finding security vulnerabilities~\cite{CM04}, resource leaks~\cite{2013_FlowDroidPreciseContext_arzt,2018_AmandroidPreciseGeneral_wei, OMJ+13, CGM16, AB16, KFSZ18, BBD+16}, enforcing policies~\cite{ARLB14}, and crypto-API misuse~\cite{RXA+19,cognicrypteclipse, KNR+17,FHM+12, EBFK13, BD16, NWA+17, ZCD+19}.
Our work studies the perspectives and beliefs of practitioners regarding SASTs through 20 in-depth interviews, and is closely related to work in three areas, namely prior studies on the usability of static analysis tools, research on evaluating SASTs, and the study of general security practices in industry.

\myparagraphnew{Usability of Static Analysis Tools}
Researchers have studied how practitioners use static analysis tools and their perspectives on improving the output of static analysis tools~\cite{APM+07,APH+08,AP08,EN08,BBC+10,DFLO19}.
In particular, Johnson \etal~\cite{JSMB13} found that poorly presented output, including false positives, is one of the main problems from developers' perspective.
This has been confirmed by later studies~\cite{CB16}, where they suggested that the false positive rate should be around or below 20\%.
In a similar vein, Distefano \etal~\cite{DFLO19} recognized that while false negatives matter, it is difficult to quantify false negative rate compared to false positive rate, and thus, it is prudent to focus on optimizing the latter.
Our work complements existing literature by detailing how practitioners across the industry choose and perceive \sasts{}~(\fnumber{4}, \fnumber{5}, \fnumber{7}).
However, our qualitative findings deviate from prior work (in the context of security), \ie we find that developers prefer low false negatives, and are willing to tolerate high false positive rates (\fnumber{10}, \fnumber{11}) provided the tool detects vulnerabilities.

Further, recent work~\cite{APH+08,AP08, SVJ+15a} proposes allowing subjective interpretations of defect warnings for productivity, \ie the notion of ``effective false positives'', which has been adapted in the context of security by Wickert \etal~\cite{WBS+22}.
We study whether this concept works in the context of security, \ie whether practitioners see merit in letting developers decide what constitutes a vulnerability, and find that it does not, drawing attention to the risks of letting developers become the arbiters of false positives (\fnumber{12}).

Finally, prior work has focused on particular usability aspects of static analysis tools, such as the use cases and their contexts~\cite{NWA20, VPP+20a,NSB22}, and filtering warning messages~\cite{IRFW19}.
Our analysis of the pain points experienced by developers echoes some of these concerns (\fnumber{17}), highlights unique challenges practitioners face (\fnumber{4}, \fnumber{5}, \fnumber{15}), and culminates in our discussion of a path forward for both researchers and practitioners towards better and more useful SASTs.

\myparagraphnew{Evaluating \sasts} Historically, security researchers have focused on creating benchmarks for evaluating \sasts with a focus on precision, recall, and efficiency, with the help of benchmarks, such as
ICC-Bench~\cite{2018_AmandroidPreciseGeneral_wei}, DroidBench~\cite{FAR+13}, CryptoAPI-Bench~\cite{ARY19}, ApacheCryptoAPIBench~\cite{AXR+22}, OWASP Benchmark~\cite{owasp-benchmark}, Parametric Crypto Misuse Benchmark~\cite{WRE+19}, Ghera~\cite{MR17}, and CamBench~\cite{SWK+22}.
Moreover, prior work has proposed automated evaluation using benchmarks~\cite{PBW18,LPP+22}.
Our study reveals that practitioners, in general, do not trust third-party benchmarks, due to them being basic (\ie not representative of real vulnerabilities), or worse, biased in favor of a specific SAST tool.
Since SAST tools often accompany custom benchmarks claimed to be general (as is the case with many of the benchmarks above), we cannot deny this perception.

A recent body of research considers the limitations of existing benchmarks and leverages mutation testing to uncover flaws in the detection capabilities of SASTs~\cite{BKM+18,AKM+21, ACK+22, AKM+21-Demo, AAR+23-demo}.
As we discuss in Section~\ref{sec:discussion}, enabling better benchmarks or automated evaluation of SASTs using such evolving approaches may be a path forward towards providing developers with what they care most, \ie SASTs that can detect real, valuable, vulnerabilities.

\myparagraphnew{Study of Security Practices in Industry}
Finally, prior work has studied the relationship between developers' security expertise, and the actual implementation of secure software.
For example, researchers have identified that practitioners need security-specific support in the form of developer-friendly APIs and supporting tools~\cite{ASW+17a, GIW+18, GALF20, NKMB16, SJM+15} to make better security choices, recognizing that practitioners may not know enough about security~\cite{GS16,Wv08,CB16}.
Furthermore, researchers have explored how developers address security-specific tasks and the challenges they face~\cite{ASW+17, GKB+22, HGK+19, SKDM22, JFB+22}.
Our work complements previous studies by exploring how developers choose \sasts (\fnumber{4}, \fnumber{5}).
Moreover, we explore how developers depend on \sasts to cover their knowledge gaps (\fnumber{6}) and the challenges, beliefs, and perceptions associated with implementing security in software with the help of \sasts~(\fnumber{8}--\fnumber{17}).

    \section{Conclusion}
\label{sec:conclusion}

This paper provides a comprehensive understanding of how practitioners with diverse business and security needs choose \sasts, and their perspectives and assumptions about limitations of \sasts.
By qualitatively analyzing the responses from $20$ in-depth interviews, we uncover $17$ key findings that demonstrate that contrary to existing literature, practitioners have a higher level of tolerance for false positives, and prioritize avoiding false negatives.
Moreover, we find that practitioners, regardless of their strong preference for security, rely on reputation to choose \sasts, as they {\em do not trust benchmarks} or find them reliable.
Finally, practitioners may be overconfident in assuming their ability to address a \sast's flaw with manual analysis, and are generally hesitant to report such flaws.
We conclude with research directions towards automated evaluation of \sasts, aligning \sasts with what developers desire, and creating dedicated protocols for reporting flaws in \sasts{}.

    \section*{Acknowledgments}
The authors have been supported by the NSF-$1815336$, NSF CNS-$2132281$, CNS-$2132285$ and CCF-$1955853$ grants, and a CoVA CCI Dissertation Fellowship.
Any opinions, findings, and conclusions herein are the authors' and do not necessarily reflect those of the sponsors.
    \bibliographystyle{IEEEtran}
    \bibliography{thematic_analysis_in_security,thematic_analysis,ethics,methodology,static-analysis,misc,developers}

\begin{thebibliography}{10}
\providecommand{\url}[1]{#1}
\csname url@samestyle\endcsname
\providecommand{\newblock}{\relax}
\providecommand{\bibinfo}[2]{#2}
\providecommand{\BIBentrySTDinterwordspacing}{\spaceskip=0pt\relax}
\providecommand{\BIBentryALTinterwordstretchfactor}{4}
\providecommand{\BIBentryALTinterwordspacing}{\spaceskip=\fontdimen2\font plus
\BIBentryALTinterwordstretchfactor\fontdimen3\font minus
  \fontdimen4\font\relax}
\providecommand{\BIBforeignlanguage}[2]{{%
\expandafter\ifx\csname l@#1\endcsname\relax
\typeout{** WARNING: IEEEtran.bst: No hyphenation pattern has been}%
\typeout{** loaded for the language `#1'. Using the pattern for}%
\typeout{** the default language instead.}%
\else
\language=\csname l@#1\endcsname
\fi
#2}}
\providecommand{\BIBdecl}{\relax}
\BIBdecl

\bibitem{solarwinds-response}
U.~S. G.~A. Office, ``{{SolarWinds Cyberattack Demands Significant Federal}}
  and {{Private-Sector Response}} (infographic) | {{U}}.{{S}}. {{GAO}},''
  https://www.gao.gov/blog/solarwinds-cyberattack-demands-significant-federal-and-private-sector-response-infographic.

\bibitem{certify-software-senate-2021}
R.~R. Torres, ``H.{{R}}.4611 - {{DHS Software Supply Chain Risk Management
  Act}} of 2021,'' Oct. 2021.

\bibitem{cyber-shield-act}
{United States Senate}, ``S.965 - cyber shield act of 2021,''
  \url{https://www.congress.gov/bill/117th-congress/senate-bill/965}, 2021.

\bibitem{ioxt23}
``{{ioXt}} - {{The Global Standard}} for {{IoT Security}},''
  https://www.ioxtalliance.org, Mar. 2023.

\bibitem{Hou21}
T.~W. House, ``Executive {{Order}} on {{Improving}} the {{Nation}}'s
  {{Cybersecurity}},''
  https://www.whitehouse.gov/briefing-room/presidential-actions/2021/05/12/executive-order-on-improving-the-nations-cybersecurity/,
  May 2021.

\bibitem{gartner-application-testing-billion-dollars}
G.~Inc, ``Application {{Security Testing Reviews}} 2023 | {{Gartner Peer
  Insights}},'' https://www.gartner.com/market/application-security-testing,
  2023.

\bibitem{github-code-scan}
``Enabling code scanning for a repository - {{GitHub Docs}},''
  https://docs.github.com/en/free-pro-team@latest/github/finding-security-vulnerabilities-and-errors-in-your-code/enabling-code-scanning-for-a-repository,
  accessed Apr, 2023.

\bibitem{ACK+22}
A.~S. Ami, N.~Cooper, K.~Kafle, K.~Moran, D.~Poshyvanyk, and A.~Nadkarni, ``Why
  {{Crypto-detectors Fail}}: {{A Systematic Evaluation}} of {{Cryptographic
  Misuse Detection Techniques}},'' in \emph{2022 {{IEEE Symposium}} on
  {{Security}} and {{Privacy}} ({{S}}\&{{P}})}.\hskip 1em plus 0.5em minus
  0.4em\relax {San Francisco, CA, USA}: {IEEE Computer Society}, May 2022, pp.
  397--414.

\bibitem{AKM+21}
A.~S. Ami, K.~Kafle, K.~Moran, A.~Nadkarni, and D.~Poshyvanyk, ``Systematic
  {{Mutation-Based Evaluation}} of the {{Soundness}} of {{Security-Focused
  Android Static Analysis Techniques}},'' \emph{ACM Transactions on Privacy and
  Security}, vol.~24, no.~3, pp. 15:1--15:37, Feb. 2021.

\bibitem{online-appendix}
``False negatives kill,''
  \url{https://github.com/Secure-Platforms-Lab-W-M/false-negatives-kill}, Apr.
  2023.

\bibitem{SnowballSamplinggoodman1961a}
L.~A. Goodman, ``Snowball {{Sampling}},'' \emph{The Annals of Mathematical
  Statistics}, vol.~32, no.~1, pp. 148--170, 1961.

\bibitem{octoverse2021}
``The {{State}} of the {{Octoverse}} | {{The State}} of the {{Octoverse}}
  explores a year of change with new deep dives into writing code faster,
  creating documentation and how we build sustainable communities on
  {{GitHub}}.'' https://octoverse.github.com/2021/.

\bibitem{coverity-scan}
``Coverity {{Scan}} - {{Projects Using Scan}},''
  https://scan.coverity.com/projects.

\bibitem{Ada15}
\BIBentryALTinterwordspacing
W.~C. Adams, ``Conducting {{Semi-Structured Interviews}},'' in \emph{Handbook
  of {{Practical Program Evaluation}}}, K.~E. Newcomer, H.~P. Hatry, and J.~S.
  Wholey, Eds.\hskip 1em plus 0.5em minus 0.4em\relax {John Wiley \& Sons,
  Inc.}, 2015, pp. 492--505. [Online]. Available:
  \url{https://onlinelibrary.wiley.com/doi/10.1002/9781119171386.ch19}
\BIBentrySTDinterwordspacing

\bibitem{HA05}
S.~Hove and B.~Anda, ``{Experiences from Conducting Semi-Structured Interviews
  in Empirical Software Engineering Research},'' in \emph{11th {{IEEE
  International Software Metrics Symposium}} ({{METRICS}}'05)}, 2005, pp. 10
  pp.--23.

\bibitem{CRM+94}
C.~Corbridge, G.~Rugg, N.~P. Major, N.~R. Shadbolt, and A.~M. Burton,
  ``Laddering: Technique and tool use in knowledge acquisition,''
  \emph{Knowledge Acquisition}, vol.~6, no.~3, pp. 315--341, Sep. 1994.

\bibitem{KD12}
\BIBentryALTinterwordspacing
E.~Kenneally and D.~Dittrich, ``The {{Menlo Report}}: {{Ethical Principles
  Guiding Information}} and {{Communication Technology Research}},''
  \emph{{SSRN} Electronic Journal}, 2012. [Online]. Available:
  \url{https://papers.ssrn.com/abstract=2445102}
\BIBentrySTDinterwordspacing

\bibitem{DKB13}
\BIBentryALTinterwordspacing
D.~Dittrich, E.~Kenneally, and M.~Bailey, ``Applying {{Ethical Principles}} to
  {{Information}} and {{Communication Technology Research}}: {{A Companion}} to
  the {{Menlo Report}},'' \emph{{SSRN} Electronic Journal}, 2013. [Online].
  Available: \url{https://papers.ssrn.com/abstract=2342036}
\BIBentrySTDinterwordspacing

\bibitem{braun2021thematic}
\BIBentryALTinterwordspacing
V.~Braun and V.~Clarke, \emph{Thematic Analysis: {{A}} Practical Guide}.\hskip
  1em plus 0.5em minus 0.4em\relax {SAGE Publications}, 2021. [Online].
  Available: \url{https://books.google.com/books?id=eMArEAAAQBAJ}
\BIBentrySTDinterwordspacing

\bibitem{FCV+21}
K.~R. Fulton, A.~Chan, D.~Votipka, M.~Hicks, and M.~L. Mazurek, ``Benefits and
  drawbacks of adopting a secure programming language: {{Rust}} as a case
  study,'' in \emph{Seventeenth Symposium on Usable Privacy and Security
  ({{SOUPS}} 2021)}.\hskip 1em plus 0.5em minus 0.4em\relax {USENIX
  Association}, Aug. 2021, pp. 597--616.

\bibitem{PTL+20}
H.~Palombo, A.~Z. Tabari, D.~Lende, J.~Ligatti, and X.~Ou, ``An ethnographic
  understanding of software ({{In}}){{Security}} and a {{Co-Creation}} model to
  improve secure software development,'' in \emph{Sixteenth Symposium on Usable
  Privacy and Security ({{SOUPS}} 2020)}.\hskip 1em plus 0.5em minus
  0.4em\relax {USENIX Association}, Aug. 2020, pp. 205--220.

\bibitem{AC19}
H.~Assal and S.~Chiasson, ``{\emph{'}}{{{\emph{Think}}}}{\emph{ secure from the
  beginning'}}: {{A Survey}} with {{Software Developers}},'' in
  \emph{Proceedings of the 2019 {{CHI Conference}} on {{Human Factors}} in
  {{Computing Systems}}}.\hskip 1em plus 0.5em minus 0.4em\relax {Glasgow
  Scotland Uk}: {ACM}, May 2019, pp. 1--13.

\bibitem{XWM14}
S.~Xiao, J.~Witschey, and E.~{Murphy-Hill}, ``Social influences on secure
  development tool adoption: Why security tools spread,'' in \emph{Proceedings
  of the 17th {{ACM}} Conference on {{Computer}} Supported Cooperative Work \&
  Social Computing}.\hskip 1em plus 0.5em minus 0.4em\relax {Baltimore Maryland
  USA}: {ACM}, Feb. 2014, pp. 1095--1106.

\bibitem{WZW+15}
J.~Witschey, O.~Zielinska, A.~Welk, E.~{Murphy-Hill}, C.~Mayhorn, and
  T.~Zimmermann, ``Quantifying developers' adoption of security tools,'' in
  \emph{Proceedings of the 2015 10th {{Joint Meeting}} on {{Foundations}} of
  {{Software Engineering}}}.\hskip 1em plus 0.5em minus 0.4em\relax {Bergamo
  Italy}: {ACM}, Aug. 2015, pp. 260--271.

\bibitem{JSMB13}
\BIBentryALTinterwordspacing
B.~Johnson, Y.~Song, E.~Murphy-Hill, and R.~Bowdidge, ``Why {{Don}}'t
  {{Software Developers Use Static Analysis Tools}} to {{Find Bugs}}?'' in
  \emph{35th {{International Conference}} on {{Software Engineering}}
  ({{ICSE}})}.\hskip 1em plus 0.5em minus 0.4em\relax {IEEE}, 2013, pp.
  672--681. [Online]. Available:
  \url{http://ieeexplore.ieee.org/document/6606613/}
\BIBentrySTDinterwordspacing

\bibitem{IRFW19}
N.~Imtiaz, A.~Rahman, E.~Farhana, and L.~Williams, ``Challenges with
  {{Responding}} to {{Static Analysis Tool Alerts}},'' in
  \emph{{{IEEE}}/{{ACM}} 16th {{International Conference}} on {{Mining Software
  Repositories}} ({{MSR}})}, 2019, pp. 245--249.

\bibitem{SAE+18}
\BIBentryALTinterwordspacing
C.~Sadowski, E.~Aftandilian, A.~Eagle, L.~Miller-Cushon, and C.~Jaspan,
  ``{Lessons from Building Static Analysis Tools at Google},'' \emph{Commun.
  ACM}, vol.~61, no.~4, p. 58–66, mar 2018. [Online]. Available:
  \url{https://doi.org/10.1145/3188720}
\BIBentrySTDinterwordspacing

\bibitem{RXA+19}
S.~Rahaman, Y.~Xiao, S.~Afrose, F.~Shaon, K.~Tian, M.~Frantz, M.~Kantarcioglu,
  and D.~D. Yao, ``{{CryptoGuard}}: {{High Precision Detection}} of
  {{Cryptographic Vulnerabilities}} in {{Massive-sized Java Projects}},'' in
  \emph{Proceedings of the 2019 {{ACM SIGSAC Conference}} on {{Computer}} and
  {{Communications Security}} - {{CCS}} '19}.\hskip 1em plus 0.5em minus
  0.4em\relax {London, United Kingdom}: {ACM Press}, 2019, pp. 2455--2472.

\bibitem{2018_AmandroidPreciseGeneral_wei}
F.~Wei, S.~Roy, X.~Ou, and {Robby}, ``Amandroid: {{A Precise}} and {{General
  Inter-component Data Flow Analysis Framework}} for {{Security Vetting}} of
  {{Android Apps}},'' \emph{ACM Transactions on Privacy and Security}, vol.~21,
  no.~3, pp. 1--32, Apr. 2018.

\bibitem{2013_FlowDroidPreciseContext_arzt}
S.~Arzt, S.~Rasthofer, C.~Fritz, E.~Bodden, A.~Bartel, J.~Klein, Y.~Le~Traon,
  D.~Octeau, and P.~McDaniel, ``{{FlowDroid}}: Precise context, flow, field,
  object-sensitive and lifecycle-aware taint analysis for {{Android}} apps,''
  in \emph{Proceedings of the 35th {{ACM SIGPLAN Conference}} on {{Programming
  Language Design}} and {{Implementation}} - {{PLDI}} '14}.\hskip 1em plus
  0.5em minus 0.4em\relax {Edinburgh, United Kingdom}: {ACM Press}, 2013, pp.
  259--269.

\bibitem{FAR+13}
C.~Fritz, S.~Arzt, S.~Rasthofer, E.~Bodden, A.~Bartel, J.~Klein, Y.~le~Traon,
  D.~Octeau, and P.~McDaniel, ``Highly precise taint analysis for android
  applications,'' {EC SPRIDE}, Tech. Rep. TUD-CS-2013-0113, May 2013.

\bibitem{KFSZ18}
W.~Klieber, L.~Flynn, W.~Snavely, and M.~Zheng, ``Practical {{Precise
  Taint-flow Static Analysis}} for {{Android App Sets}},'' in \emph{Proceedings
  of the 13th {{International Conference}} on {{Availability}}, {{Reliability}}
  and {{Security}}}.\hskip 1em plus 0.5em minus 0.4em\relax {Hamburg Germany}:
  {ACM}, Aug. 2018, pp. 1--7.

\bibitem{2015_ScalablePreciseTaint_huang}
W.~Huang, Y.~Dong, A.~Milanova, and J.~Dolby, ``Scalable and precise taint
  analysis for {{Android}},'' in \emph{Proceedings of the 2015 {{International
  Symposium}} on {{Software Testing}} and {{Analysis}} - {{ISSTA}} 2015}.\hskip
  1em plus 0.5em minus 0.4em\relax {Baltimore, MD, USA}: {ACM Press}, 2015, pp.
  106--117.

\bibitem{AB16}
S.~Arzt and E.~Bodden, ``{{StubDroid}}: {{Automatic}} inference of precise
  data-flow summaries for the android framework,'' in \emph{International
  Conference for Software Engineering ({{ICSE}})}, May 2016.

\bibitem{SFZ11}
H.~Shen, J.~Fang, and J.~Zhao, ``{{EFindBugs}}: {{Effective Error Ranking}} for
  {{FindBugs}},'' in \emph{Verification and {{Validation}} 2011 {{Fourth IEEE
  International Conference}} on {{Software Testing}}}, Mar. 2011, pp. 299--308.

\bibitem{CB16}
M.~Christakis and C.~Bird, ``What developers want and need from program
  analysis: An empirical study,'' in \emph{Proceedings of the 31st
  {{IEEE}}/{{ACM International Conference}} on {{Automated Software
  Engineering}}}.\hskip 1em plus 0.5em minus 0.4em\relax {Singapore Singapore}:
  {ACM}, Aug. 2016, pp. 332--343.

\bibitem{BBC+10}
A.~Bessey, K.~Block, B.~Chelf, A.~Chou, B.~Fulton, S.~Hallem, C.~{Henri-Gros},
  A.~Kamsky, S.~McPeak, and D.~Engler, ``A {{Few Billion Lines}} of {{Code
  Later}}: {{Using Static Analysis}} to {{Find Bugs}} in the {{Real World}},''
  \emph{Communications of the ACM}, vol.~53, no.~2, pp. 66--75, Feb. 2010.

\bibitem{SVJ+15a}
\BIBentryALTinterwordspacing
C.~Sadowski, J.~Van~Gogh, C.~Jaspan, E.~Soderberg, and C.~Winter, ``Tricorder:
  {{Building}} a {{Program Analysis Ecosystem}},'' in \emph{2015
  {{IEEE}}/{{ACM}} 37th {{IEEE International Conference}} on {{Software
  Engineering}}}.\hskip 1em plus 0.5em minus 0.4em\relax {IEEE}, 2015, pp.
  598--608. [Online]. Available:
  \url{https://ieeexplore.ieee.org/document/7194609/}
\BIBentrySTDinterwordspacing

\bibitem{WBS+22}
A.-K. Wickert, L.~Baumg{\"a}rtner, M.~Schlichtig, K.~Narasimhan, and M.~Mezini,
  ``To {{Fix}} or {{Not}} to {{Fix}}: {{A Critical Study}} of
  {{Crypto-misuses}} in the {{Wild}},'' in \emph{The 21th {{IEEE International
  Conference}} on {{Trust}}, {{Security}} and {{Privacy}} in {{Computing}} and
  {{Communications}} ({{TrustCom}})}.\hskip 1em plus 0.5em minus 0.4em\relax
  {IEEE}, 2022.

\bibitem{Wv08}
G.~Wurster and P.~C. {van Oorschot}, ``The developer is the enemy,'' in
  \emph{Proceedings of the 2008 {{New Security Paradigms Workshop}}}.\hskip 1em
  plus 0.5em minus 0.4em\relax {Lake Tahoe California USA}: {ACM}, Sep. 2008,
  pp. 89--97.

\bibitem{AB96}
D.~{Austen-Smith} and J.~S. Banks, ``Information {{Aggregation}},
  {{Rationality}}, and the {{Condorcet Jury Theorem}},'' \emph{The American
  Political Science Review}, vol.~90, no.~1, pp. 34--45, 1996.

\bibitem{DLB05}
H.~Dorussen, H.~Lenz, and S.~Blavoukos, ``Assessing the {{Reliability}} and
  {{Validity}} of {{Expert Interviews}},'' \emph{European Union Politics},
  vol.~6, no.~3, pp. 315--337, Sep. 2005.

\bibitem{github-actions}
``Features \textbullet{} {{GitHub Actions}},''
  https://github.com/features/actions, accessed Apr, 2023.

\bibitem{SARIF}
O.~S. ~, ``Static analysis results interchange format ({{SARIF}}) version
  2.1.0,'' https://docs.oasis-open.org/sarif/sarif/v2.1.0/sarif-v2.1.0.html,
  Mar. 2020.

\bibitem{CM04}
B.~Chess and G.~McGraw, ``Static analysis for security,'' \emph{IEEE Security
  Privacy}, vol.~2, no.~6, pp. 76--79, Nov. 2004.

\bibitem{OMJ+13}
D.~Octeau, P.~McDaniel, S.~Jha, A.~Bartel, E.~Bodden, J.~Klein, and Y.~L.
  Traon, ``Effective {{Inter-Component}} communication mapping in android:
  {{An}} essential step towards holistic security analysis,'' in \emph{22nd
  {{USENIX}} Security Symposium ({{USENIX}} Security 13)}.\hskip 1em plus 0.5em
  minus 0.4em\relax {Washington, D.C.}: {USENIX Association}, Aug. 2013, pp.
  543--558.

\bibitem{CGM16}
S.~Calzavara, I.~Grishchenko, and M.~Maffei, ``{{HornDroid}}: {{Practical}} and
  {{Sound Static Analysis}} of {{Android Applications}} by {{SMT Solving}},''
  in \emph{2016 {{IEEE European Symposium}} on {{Security}} and {{Privacy}}
  ({{EuroS P}})}, Mar. 2016, pp. 47--62.

\bibitem{BBD+16}
M.~Backes, S.~Bugiel, E.~Derr, S.~Gerling, and C.~Hammer, ``R-{{Droid}}:
  {{Leveraging Android App Analysis}} with {{Static Slice Optimization}},'' in
  \emph{Proceedings of the 11th {{ACM}} on {{Asia Conference}} on {{Computer}}
  and {{Communications Security}}}, ser. {{ASIA CCS}} '16.\hskip 1em plus 0.5em
  minus 0.4em\relax {New York, NY, USA}: {ACM}, 2016, pp. 129--140.

\bibitem{ARLB14}
S.~Arzt, S.~Rasthofer, E.~Lovat, and E.~Bodden, ``{{DroidForce}}: {{Enforcing}}
  complex, data-centric, system-wide policies in android,'' in
  \emph{International Conference on Availability, Reliability and Security
  ({{ARES}} 2014)}.\hskip 1em plus 0.5em minus 0.4em\relax {IEEE}, Sep. 2014,
  pp. 40--49.

\bibitem{cognicrypteclipse}
``{{CogniCrypt}} - {{Secure Integration}} of {{Cryptographic Software}} |
  {{CogniCrypt}},'' Jun. 2020, \url{https://www.eclipse.org/cognicrypt/}.

\bibitem{KNR+17}
S.~Kr{\"u}ger, S.~Nadi, M.~Reif, K.~Ali, M.~Mezini, E.~Bodden, F.~G{\"o}pfert,
  F.~G{\"u}nther, C.~Weinert, D.~Demmler, and R.~Kamath, ``{{CogniCrypt}}:
  {{Supporting Developers}} in {{Using Cryptography}},'' in \emph{Proceedings
  of the {{32Nd IEEE}}/{{ACM International Conference}} on {{Automated Software
  Engineering}}}, ser. {{ASE}} 2017.\hskip 1em plus 0.5em minus 0.4em\relax
  {Piscataway, NJ, USA}: {IEEE Press}, 2017, pp. 931--936.

\bibitem{FHM+12}
S.~Fahl, M.~Harbach, T.~Muders, L.~Baumg{\"a}rtner, B.~Freisleben, and
  M.~Smith, ``Why {{Eve}} and {{Mallory Love Android}}: {{An Analysis}} of
  {{Android SSL}} (in){{Security}},'' in \emph{Proceedings of the 2012 {{ACM
  Conference}} on {{Computer}} and {{Communications Security}}}, ser. {{CCS}}
  '12.\hskip 1em plus 0.5em minus 0.4em\relax {New York, NY, USA}: {Association
  for Computing Machinery}, 2012, pp. 50--61.

\bibitem{EBFK13}
M.~Egele, D.~Brumley, Y.~Fratantonio, and C.~Kruegel, ``An empirical study of
  cryptographic misuse in android applications,'' in \emph{Proceedings of the
  2013 {{ACM SIGSAC}} Conference on {{Computer}} \& Communications Security -
  {{CCS}} '13}.\hskip 1em plus 0.5em minus 0.4em\relax {Berlin, Germany}: {ACM
  Press}, 2013, pp. 73--84.

\bibitem{BD16}
A.~Braga and R.~Dahab, ``Mining {{Cryptography Misuse}} in {{Online Forums}},''
  in \emph{2016 {{IEEE International Conference}} on {{Software Quality}},
  {{Reliability}} and {{Security Companion}} ({{QRS-C}})}, Aug. 2016, pp.
  143--150.

\bibitem{NWA+17}
D.~C. Nguyen, D.~Wermke, Y.~Acar, M.~Backes, C.~Weir, and S.~Fahl, ``A
  {{Stitch}} in {{Time}}: {{Supporting Android Developers}} in {{WritingSecure
  Code}},'' in \emph{Proceedings of the 2017 {{ACM SIGSAC Conference}} on
  {{Computer}} and {{Communications Security}}}, ser. {{CCS}} '17.\hskip 1em
  plus 0.5em minus 0.4em\relax {New York, NY, USA}: {Association for Computing
  Machinery}, Oct. 2017, pp. 1065--1077.

\bibitem{ZCD+19}
L.~Zhang, J.~Chen, W.~Diao, S.~Guo, J.~Weng, and K.~Zhang, ``{{CryptoREX}}:
  Large-scale analysis of cryptographic misuse in {{IoT}} devices,'' in
  \emph{22nd International Symposium on Research in Attacks, Intrusions and
  Defenses ({{RAID}} 2019)}.\hskip 1em plus 0.5em minus 0.4em\relax {Chaoyang
  District, Beijing}: {USENIX Association}, Sep. 2019, pp. 151--164.

\bibitem{APM+07}
N.~Ayewah, W.~Pugh, J.~D. Morgenthaler, J.~Penix, and Y.~Zhou, ``Evaluating
  static analysis defect warnings on production software,'' in
  \emph{Proceedings of the 7th {{ACM SIGPLAN-SIGSOFT}} Workshop on {{Program}}
  Analysis for Software Tools and Engineering - {{PASTE}} '07}.\hskip 1em plus
  0.5em minus 0.4em\relax {San Diego, California, USA}: {ACM Press}, 2007, pp.
  1--8.

\bibitem{APH+08}
N.~Ayewah, W.~Pugh, D.~Hovemeyer, J.~D. Morgenthaler, and J.~Penix, ``Using
  {{Static Analysis}} to {{Find Bugs}},'' \emph{IEEE Software}, vol.~25, no.~5,
  pp. 22--29, Sep. 2008.

\bibitem{AP08}
N.~Ayewah and W.~Pugh, ``A {{Report}} on a {{Survey}} and {{Study}} of {{Static
  Analysis Users}},'' in \emph{Proceedings of the 2008 Workshop on {{Defects}}
  in Large Software Systems - {{DEFECTS}} '08}.\hskip 1em plus 0.5em minus
  0.4em\relax {Seattle, Washington}: {ACM Press}, 2008, p.~1.

\bibitem{EN08}
P.~Emanuelsson and U.~Nilsson, ``A {{Comparative Study}} of {{Industrial Static
  Analysis Tools}},'' \emph{Electronic Notes in Theoretical Computer Science},
  vol. 217, pp. 5--21, Jul. 2008.

\bibitem{DFLO19}
D.~Distefano, M.~F{\"a}hndrich, F.~Logozzo, and P.~W. O'Hearn, ``Scaling static
  analyses at {{Facebook}},'' \emph{Communications of the ACM}, vol.~62, no.~8,
  pp. 62--70, Jul. 2019.

\bibitem{NWA20}
L.~Nguyen Quang~Do, J.~Wright, and K.~Ali, ``Why {{Do Software Developers Use
  Static Analysis Tools}}? {{A User-Centered Study}} of {{Developer Needs}} and
  {{Motivations}},'' \emph{IEEE Transactions on Software Engineering}, vol.~48,
  no.~3, pp. 835--847, 2020.

\bibitem{VPP+20a}
C.~Vassallo, S.~Panichella, F.~Palomba, S.~Proksch, H.~C. Gall, and A.~Zaidman,
  ``How developers engage with static analysis tools in different contexts,''
  \emph{Empirical Software Engineering}, vol.~25, no.~2, pp. 1419--1457, Mar.
  2020.

\bibitem{NSB22}
M.~Nachtigall, M.~Schlichtig, and E.~Bodden, ``A large-scale study of usability
  criteria addressed by static analysis tools,'' in \emph{Proceedings of the
  31st {{ACM SIGSOFT International Symposium}} on {{Software Testing}} and
  {{Analysis}}}.\hskip 1em plus 0.5em minus 0.4em\relax {Virtual South Korea}:
  {ACM}, Jul. 2022, pp. 532--543.

\bibitem{ARY19}
S.~Afrose, S.~Rahaman, and D.~Yao, ``{{CryptoAPI-Bench}}: {{A}} comprehensive
  benchmark on java cryptographic {{API}} misuses,'' in \emph{2019 {{IEEE}}
  Cybersecurity Development ({{SecDev}})}, Sep. 2019, pp. 49--61.

\bibitem{AXR+22}
S.~Afrose, Y.~Xiao, S.~Rahaman, B.~Miller, and D.~D. Yao, ``Evaluation of
  {{Static Vulnerability Detection Tools}} with {{Java Cryptographic API
  Benchmarks}},'' \emph{IEEE Transactions on Software Engineering}, 2022.

\bibitem{owasp-benchmark}
``{{OWASP Benchmark}} | {{OWASP Foundation}},''
  https://owasp.org/www-project-benchmark/.

\bibitem{WRE+19}
A.-K. Wickert, M.~Reif, M.~Eichberg, A.~Dodhy, and M.~Mezini, ``A {{Dataset}}
  of {{Parametric Cryptographic Misuses}},'' in \emph{2019 {{IEEE}}/{{ACM}}
  16th {{International Conference}} on {{Mining Software Repositories}}
  ({{MSR}})}, May 2019, pp. 96--100.

\bibitem{MR17}
J.~Mitra and V.-P. Ranganath, ``Ghera: {{A Repository}} of {{Android App
  Vulnerability Benchmarks}},'' in \emph{Proceedings of the 13th
  {{International Conference}} on {{Predictive Models}} and {{Data Analytics}}
  in {{Software Engineering}}}.\hskip 1em plus 0.5em minus 0.4em\relax {Toronto
  Canada}: {ACM}, Nov. 2017, pp. 43--52.

\bibitem{SWK+22}
M.~Schlichtig, A.-K. Wickert, S.~Kr{\"u}ger, E.~Bodden, and M.~Mezini,
  ``{{CamBench}} -- {{Cryptographic API Misuse Detection Tool Benchmark
  Suite}},'' Apr. 2022.

\bibitem{PBW18}
F.~Pauck, E.~Bodden, and H.~Wehrheim, ``Do {{Android Taint Analysis Tools Keep
  Their Promises}}?'' in \emph{Proceedings of the 2018 26th {{ACM Joint
  Meeting}} on {{European Software Engineering Conference}} and {{Symposium}}
  on the {{Foundations}} of {{Software Engineering}}}, ser. {{ESEC}}/{{FSE}}
  2018.\hskip 1em plus 0.5em minus 0.4em\relax {New York, NY, USA}: {ACM},
  2018, pp. 331--341.

\bibitem{LPP+22}
L.~Luo, F.~Pauck, G.~Piskachev, M.~Benz, I.~Pashchenko, M.~Mory, E.~Bodden,
  B.~Hermann, and F.~Massacci, ``{{TaintBench}}: {{Automatic}} real-world
  malware benchmarking of {{Android}} taint analyses,'' \emph{Empirical
  Software Engineering}, vol.~27, no.~1, p.~16, Jan. 2022.

\bibitem{BKM+18}
R.~Bonett, K.~Kafle, K.~Moran, A.~Nadkarni, and D.~Poshyvanyk, ``Discovering
  {{Flaws}} in {{Security-Focused Static Analysis Tools}} for {{Android}} using
  {{Systematic Mutation}},'' in \emph{27th {{USENIX Security Symposium}}
  ({{USENIX Security}} 18)}.\hskip 1em plus 0.5em minus 0.4em\relax {Baltimore,
  MD}: {USENIX Association}, Aug. 2018, pp. 1263--1280.

\bibitem{AKM+21-Demo}
A.~S. Ami, K.~Kafle, K.~Moran, A.~Nadkarni, and D.~Poshyvanyk, ``{Demo:
  Mutation-based Evaluation of Security-focused Static Analysis Tools for
  Android.}'' in \emph{Proceedings of the 43rd IEEE/ACM International
  Conference on Software Engineering (ICSE'21), Formal Tool Demonstration
  Track}, May 2021.

\bibitem{AAR+23-demo}
A.~S. Ami, S.~Y. Ahmed, R.~M. Redoy, N.~Cooper, K.~Kafle, K.~Moran,
  A.~Nadkarni, and D.~Poshyvanyk, ``{MASC: A Tool for Mutation-based Evaluation
  of Static Crypto-API Misuse Detectors},'' in \emph{Proceedings of the 31st
  ACM Joint European Software Engineering Conference and Symposium on the
  Foundations of Software Engineering}, ser. ESEC/FSE 2023.\hskip 1em plus
  0.5em minus 0.4em\relax New York, NY, USA: Association for Computing
  Machinery, Dec. 2023.

\bibitem{ASW+17a}
Y.~Acar, C.~Stransky, D.~Wermke, C.~Weir, M.~L. Mazurek, and S.~Fahl,
  ``Developers {{Need Support}}, {{Too}}: {{A Survey}} of {{Security Advice}}
  for {{Software Developers}},'' in \emph{2017 {{IEEE Cybersecurity
  Development}} ({{SecDev}})}, Sep. 2017, pp. 22--26.

\bibitem{GIW+18}
P.~L. Gorski, L.~L. Iacono, D.~Wermke, C.~Stransky, S.~M{\"o}ller, Y.~Acar, and
  S.~Fahl, ``Developers {{Deserve Security Warnings}}, {{Too}}: {{On}} the
  {{Effect}} of {{Integrated Security Advice}} on {{Cryptographic API
  Misuse}},'' in \emph{Fourteenth {{Symposium}} on {{Usable Privacy}} and
  {{Security}}, {{SOUPS}} 2018, {{Baltimore}}, {{MD}}, {{USA}}, {{August}}
  12-14, 2018}, M.~E. Zurko and H.~R. Lipford, Eds.\hskip 1em plus 0.5em minus
  0.4em\relax {USENIX Association}, 2018, pp. 265--281.

\bibitem{GALF20}
P.~L. Gorski, Y.~Acar, L.~Lo~Iacono, and S.~Fahl, ``Listen to {{Developers}}!
  {{A Participatory Design Study}} on {{Security Warnings}} for {{Cryptographic
  APIs}},'' in \emph{Proceedings of the 2020 {{CHI Conference}} on {{Human
  Factors}} in {{Computing Systems}}}.\hskip 1em plus 0.5em minus 0.4em\relax
  {Honolulu HI USA}: {ACM}, Apr. 2020, pp. 1--13.

\bibitem{NKMB16}
S.~Nadi, S.~Kr{\"u}ger, M.~Mezini, and E.~Bodden, ``Jumping {{Through Hoops}}:
  {{Why Do Java Developers Struggle}} with {{Cryptography APIs}}?'' in
  \emph{Proceedings of the 38th {{International Conference}} on {{Software
  Engineering}}}, ser. {{ICSE}} '16.\hskip 1em plus 0.5em minus 0.4em\relax
  {New York, NY, USA}: {ACM}, 2016, pp. 935--946.

\bibitem{SJM+15}
J.~Smith, B.~Johnson, E.~{Murphy-Hill}, B.~Chu, and H.~R. Lipford, ``Questions
  developers ask while diagnosing potential security vulnerabilities with
  static analysis,'' in \emph{Proceedings of the 2015 10th {{Joint Meeting}} on
  {{Foundations}} of {{Software Engineering}}}.\hskip 1em plus 0.5em minus
  0.4em\relax {Bergamo Italy}: {ACM}, Aug. 2015, pp. 248--259.

\bibitem{GS16}
M.~Green and M.~Smith, ``Developers are {{Not}} the {{Enemy}}!: {{The Need}}
  for {{Usable Security APIs}},'' \emph{IEEE Security \& Privacy}, vol.~14,
  no.~5, pp. 40--46, Sep. 2016.

\bibitem{ASW+17}
Y.~Acar, C.~Stransky, D.~Wermke, M.~L. Mazurek, and S.~Fahl, ``Security
  developer studies with {{GitHub}} users: {{Exploring}} a convenience
  sample,'' in \emph{Thirteenth Symposium on Usable Privacy and Security
  ({{SOUPS}} 2017)}.\hskip 1em plus 0.5em minus 0.4em\relax {Santa Clara, CA}:
  {USENIX Association}, Jul. 2017, pp. 81--95.

\bibitem{GKB+22}
M.~Gutfleisch, J.~H. Klemmer, N.~Busch, Y.~Acar, M.~A. Sasse, and S.~Fahl,
  ``How {{Does Usable Security}} ({{Not}}) {{End Up}} in {{Software Products}}?
  {{Results From}} a {{Qualitative Interview Study}},'' in \emph{2022 {{IEEE
  Symposium}} on {{Security}} and {{Privacy}} ({{SP}})}, May 2022, pp.
  893--910.

\bibitem{HGK+19}
M.~Hazhirpasand, M.~Ghafari, S.~Kr{\"u}ger, E.~Bodden, and O.~Nierstrasz, ``The
  {{Impact}} of {{Developer Experience}} in {{Using Java Cryptography}},'' in
  \emph{2019 {{ACM}}/{{IEEE International Symposium}} on {{Empirical Software
  Engineering}} and {{Measurement}} ({{ESEM}})}, Sep. 2019, pp. 1--6.

\bibitem{SKDM22}
R.~Stevens, F.~B. Kokulu, A.~Doupé, and M.~L. Mazurek, ``Above and {{Beyond}}:
  {{Organizational Efforts}} to {{Complement U}}.{{S}}. {{Digital Security
  Compliance Mandates}},'' in \emph{Proceedings 2022 {{Network}} and
  {{Distributed System Security Symposium}}}.\hskip 1em plus 0.5em minus
  0.4em\relax {Internet Society}, 2022.

\bibitem{JFB+22}
J.~Jancar, M.~Fourn{\'e}, D.~D.~A. Braga, M.~Sabt, P.~Schwabe, G.~Barthe, P.-A.
  Fouque, and Y.~Acar, ````{{They}}'re not that hard to mitigate'': {{What
  Cryptographic Library Developers Think About Timing Attacks}},'' in
  \emph{2022 {{IEEE Symposium}} on {{Security}} and {{Privacy}} ({{SP}})}, May
  2022, pp. 632--649.

\bibitem{EBW22}
M.~Endres, K.~Boehnke, and W.~Weimer, ``Hashing it out: A survey of
  programmers' cannabis usage, perception, and motivation,'' in
  \emph{Proceedings of the 44th {{International Conference}} on {{Software
  Engineering}}}.\hskip 1em plus 0.5em minus 0.4em\relax {Pittsburgh
  Pennsylvania}: {ACM}, May 2022, pp. 1107--1119.

\end{thebibliography}

    \appendices
    \section{Survey Protocol}\label{app:online-survey}
To understand how practitioners perceive security tools, and whether
security is prioritized by individuals and organizations similarly, we
prepared an online survey questionnaire (questionnaire in the
online appendix~\cite{online-appendix}) and drafted a research protocol.
We piloted the initial survey with five participants. Three of the participants were graduate students, and the rest had doctoral degrees. All pilots were from computer science background, with additional experience in software engineering and/or security.
By incorporating their feedback, we improved the survey by modifications and additional descriptions as necessary.
Our final survey protocol received the approval of our Institutional Review Boards (IRBs).
The experimental protocol of both our survey and interviews included a consent form which emphasized that the data of the participants will remain confidential and de-identified.
Furthermore, a participant could optionally submit their email address to have the chance of winning one of two $\$50.00$ gift cards or the equivalent value in local currency vouchers. The winners would be chosen from qualified participants who completed the survey and provided valid responses in the survey.

\subsection{Survey Recruitment}
\label{app:ethical-oss-recruitment}
To diversify our recruitment approach in terms of experience, culture and industry contexts, we leveraged multiple recruitment channels.
We sent invitation emails describing the goal of the survey (\ie in order to learn about their professional experiences and opinions about \sasts) to our professional networks, relying on snowball sampling for recruitment, as well as to OSS developers (as previously described in Section~\ref{sec:survey-protocol-results}).

\myparagraphnew{Ethical Considerations in Recruiting OSS developers}
We collected publicly available email addresses only, and explicitly stated our recruitment procedure in our initial contact, which is common in other recent studies (\eg Endres et al.~\cite{EBW22}).
We considered several \textit{potential trade-offs} that factored into this recruitment strategy, in addition to following the guidance provided by our IRB:
$(a)$~It is \textit{difficult} to recruit practitioners across borders who have the relevant experience, \ie configured and used automated security analysis tools,
$(b)$~we were collecting publicly available information and not amplifying the visibility of the individuals' email address, and
$(c)$~we carefully considered the Menlo Report's ethical guidelines~\cite{KD12,DKB13}.
Specifically based on these guidelines, the only potential \textit{harm} to an invited person would be receiving one unsolicited email, whereas the potential benefit of this research is in helping create more secure software, for everyone, by understanding the needs and challenges of practitioners related to security analysis techniques.

\subsection{Online Survey and Data Analysis}
Our survey (provided in the online appendix~\cite{online-appendix}) consisted of Likert Scale based questions, with optional, open-ended response to clarify their selected choice(s).
Our analysis prioritized the text-based responses since these provided additional context for the selected choice(s) in Likert scale.
One of the authors open-coded the responses for analysis.
The responses of the survey, which we summarize next, guided our interview protocol.

\section{Survey Results}\label{app:survey-results}
The results of our survey helped us refine the semi-structure guide of questions for the interview.
We now describe the demographics as well as general results elicited from the survey responses.

\myparagraphnew{Demographics} Of the $39$ responses we received, $25$ worked in a full time employment, and $12$ worked as both freelancers and full-time employees.
Almost all of them ($85\%$) identified themselves as developers with $25$ of them having more than five years of professional experience and six with at least three years of experience.
$53\%$ participants helped release a new version of software or service at least on a monthly basis in the past two years, with $26\%$ on quarterly basis. All the participants ranked themselves as at least slightly knowledgeable in security, with five being extremely knowledgeable, eight very knowledgeable and $20$ moderately knowledgeable.
$50\%$ of the participants entered their location as Asia, with the rest distributed equally between North America, Europe, United Kingdom and Africa.

\myparagraphnew{Prioritizing Security by Organizations and Individuals}
Through the survey, we asked the participants to rate the importance of privacy, security against malicious attacks, ease of use, multi-platform compatibility, multitude of features and responsiveness with respect to applications they help develop from their individual perspective.
Furthermore, we asked the participants to rate how these are prioritized by their organizations based on their personal experience.
All participants individually expressed that securing against malicious attack is very important, with $83\%$ working in organizations expressing that it is of extreme importance.
However, from their organization's perspective, only $30/35$ participants shared that securing against malicious attacks is \textit{at least} very important, with two selecting slightly important and three moderately important. The remaining two participants chose not to answer.
In other words, the importance of security against malicious attacks might not be prioritized similarly by an organization and an individual of the same organization.
For similar questions about protecting privacy in software and or services, $25$ participants expressed that it is at least very important, with two selecting moderately important.
Similar to the trend observed for securing against malicious attacks, participants expressed that they think their organizations prioritizes privacy differently compared to themselves.

To summarize, {\em an organization and its practitioners can have significantly different priorities on security and privacy} for their software or services.

\myparagraphnew{Reliance on Automated and/or Manual Analysis Techniques} When asked how the participants relied on automated and manual techniques for finding security vulnerabilities,
seven participants expressed that they rely on automated techniques for reasons such as lack of security-related expertise, manual testing being time-consuming and for automatically preventing intruders from attacking their systems.
All the participants ($26/39$) who chose both automated analysis and manual analysis techniques expressed that they do it because of additional coverage, with the manual technique being used to cover corner cases, application specific logic, or out of scope issues.
Finally, the participants ($6/39$) who expressed that they rely only on manual analysis techniques shared that it is due to lack of effectiveness, or lack of resources, or due to simply being more comfortable with manual analysis techniques.

To summarize, {\em practitioners mostly rely on a combination of automated and manual techniques to increase coverage, with the only exceptions being an exclusive reliance on automated techniques due to lack of security expertise, and on only manual techniques due to expertise/comfort with the same.}

\myparagraphnew{Impact due to Unsound \sasts}
We asked participants how their software or service would get impacted in case there was a soundness issue  \sasts they use.
Interestingly, almost all practitioners expressed that \textit{even in the case of flaws of \sasts, their applications would be moderately impacted at most}, explaining that they do not entirely depend on these tools for ensuring security and instead \textit{rely on multiple tools and/or manual reviews}.

The few participants who shared that they would be significantly affected were either involved with tool development, or were entirely dependent on \sasts{}. In other words, {\em practitioners take the impact of flaws in security tools lightly as they use multiple tools and/or manual analysis techniques to overcome limitations}.

\begin{table*}[tbp]
	\centering
	\caption{Abridged version of the Semi-structured Interview Guide}
	\vspace{-1em}
	\label{tbl:interview-questions}
	\def\arraystretch{1.2}
	\begin{tabularx}{\textwidth}{p{\textwidth}}
		\toprule
		\multicolumn{1}{l}{\textit{Section A: \interviewSectionParticipants}}                                                                                                                                                                                                                              \\
		\midrule
		To get started, can you tell us about the type or domain of software you primarily develop?                                                                                                                                                                                                        \\
		How would you describe your target client for software? Is it general people, government, or other software firms?                                                                                                                                                                                 \\
		How did you get to learn about security? Does your company arrange training/workshops for you? Or self-learning?                                                                                                                                                                                   \\
		What is important in terms of security in terms of your product/software?                                                                                                                                                                                                                          \\
		What potential threats do you consider that may compromise the security of system?                                                                                                                                                                                                                 \\
		Are these the threat assumptions you normally consider in the domain you work on/at work?                                                                                                                                                                                                          \\
		\midrule
		\multicolumn{1}{l}{\textit{Section B: \interviewSectionSecAndOrg}}                                                                                                                                                                                                                                \\
		\midrule
		Do you remember being constrained by any factors, such as Deadline/Time, Requested Features, Dependencies, or others, when programming that may have affected/compromised security guarantees?                                                                                                     \\
		How would you describe the software development process you follow?                                                                                                                                                                                                                                \\
		Do you remember your organization's existing coding standards, national regulations or any other software development process aspects having any influence on security guarantees?                                                                                                                 \\
		Have you implemented security functionalities through in-house development instead of relying on third-party libraries? What situation necessiated this?                                                                                                                                           \\
		Between using third-party libraries for security-sensitive functionalities and implementing security-specific features on your own, which one do you prefer?                                                                                                                                       \\
		Can you tell us more about the team you work with?                                                                                                                                                                                                                                                 \\
		Can you tell us about your team structure and security specific functions/components?                                                                                                                                                                                                              \\
		Do you write test cases specifically for covering/testing security-related requirements? Can you give us an example?                                                                                                                                                                               \\
		What are the consequences if the security requirements in your software are not met?                                                                                                                                                                                                               \\
		\midrule
		\multicolumn{1}{l}{\textit{Section C: \interviewSectionOrgOfSast}}                                                                                                                                                                                                                               \\
		\midrule
		Why do you favor using <\sast(s)> in your organization? What events led to this decision?                                                                                                                                                                                                                       \\
		Where does the \sast come in the Software Development Life Cycle (SDLC) you follow? Can you walk us through the process?                                                                                                                                                                           \\
		Do all (security-related) team members know/receive training about the \sasts that you use?                                                                                                                                                                                                        \\
		Are there generally any \sast-specific requirements from the user/customer?                                                                                                                                                                                                                        \\
		Can you please walk us through the process of selecting such a security focused tool?                                                                                                                                                                                                              \\
		How are \sasts Security helpful for Agile/Scrum processes?                                                                                                                                                                                                                                         \\
		Can you tell us more about the events which influenced you in becoming a \sast user instead of focusing on being a manual technique based user?                                                                                                                                                    \\
		How much is generally the cost in dollar value for licensing and/or using \sasts?                                                                                                                                                                                                                  \\
		You have mentioned that your organization relies more on the \sasts over Manual Techniques (or vice versa). Why is that?                                                                                                                                                                           \\
		How do you generally handle security bugs in product?                                                                                                                                                                                                                                              \\
		\midrule
		\multicolumn{1}{l}{\textit{Section D: \interviewSectionLimitExpectSast}}                                                                                                                                                                                                                          \\
		\midrule
		Consider the following statement: \textit{"When using an automated code review/scanner, a static tool should be capable of reporting all the issues in the code as far as static analysis allows"}. What is your opinion regarding this statement?                                                   \\
		\textit{"When using an automated code review/scanner, a static tool should only show results it is 100\% certain about, even if it means it may miss a few potential issues"}, What is your opinion regarding this statement?                                                                        \\
		Depending on the difficulty/nature of vulnerable code issues, some issues might be more difficult than others to detect by tools. Therefore, would you consider tools that are not perfect (\ie may miss some vulnerabilities) to still be acceptable to use? What is your opinion regarding this? \\
		You mentioned that you prefer FN/FP over FN/FP; Do you think this is purely because of the kind of software you work on, or do you think it is shared in the general developer community or in \angled{type} developer community?                                                                  \\
		How would you describe your overall impression when using security analysis tools for analyzing custom implemented security features?                                                                                                                                                              \\
		Does the tool clearly present detected security vulnerabilities, provide any explanations, link detected security vulnerabilities to known examples? Anything else you prefer these tools should have/report that is currently not available?                                      \\
		\midrule
		\multicolumn{1}{l}{\textit{Section E: \interviewSectionImpactUnsoundSast}}                                                                                                                                                                                                                         \\
		\midrule
		Have you ever been in a situation where there was a vulnerability in your software, which should've been detected by a \sast but was not? How did you handle it?                                                                                                                                   \\
		If in case your software has a security issue which was not found due to buggy \sast, how do you handle the consequences?                                                                                                                                                                          \\
		\textit{"Just because a tool report states that there are no security errors does not mean the software is secure, since the tool itself may be buggy"}. Can you please elaborate on your opinion regarding this statement?                                                                                      \\
		Do you expect the \sast to catch everything?                                                                                                                                                                                                                                                       \\
		What happens if you find something that a \sast should catch, but does not? Do you report it to the \sast developers?                                                                                                                                                                              \\
		Have you encountered any situation where any developer tried to evade \sast security checks by abusing flaws?                                                                                                                                                                                           \\
		If you ever reported a problem to \sast developers, did you ever get a response and a follow-up fix to the issue that you reported (with example)?                                                                                                                     \\
		(Previous context) What role do you think fuzzing tools play in comparison to \sasts here? Do you think fuzzing tools can replace \sasts?                                                                                                                                                          \\
		\midrule
		\multicolumn{1}{l}{\textit{Section F: \interviewSectionChallengeSolution}}                                                                                                                                                                                                                        \\
		\midrule
		Have you ever considered designing and using an in-house \sast? What limitations of existing tools motivated you to do so?                                                                                                                                                                         \\
		If you were given unlimited resources to fix/create the perfect \sast, what issue would you address before anything else?                                                                                                                                                                          \\
		Do you have any kind of final thoughts or anything that you would like to follow up on?                                                                                                                                                                                                               \\
		\bottomrule
	\end{tabularx}
\end{table*}

    \newpage
    \newpage
\section{Meta-Review}\label{sec:meta}
\subsection{Summary}
This paper presents a qualitative study to understand the industry's viewpoint on program analysis based security testing tools, to explore organizations' selection criteria of \sast{} tools, and to understand how practitioners understand and work around the limitations of these tools.

\subsection{Scientific Contributions}
\begin{itemize}
    \item Provides a Valuable Step Forward in an Established Field
    \item Independent Confirmation of Important Results with Limited Prior Research
    \item Addresses a Long-Known Issue
    \item Establishes a New Research Direction
\end{itemize}

\subsection{Reasons for Acceptance}
\begin{enumerate}
\item This work provides valuable qualitative understanding of the industry needs and use of SASTs and the organizational barriers that prevent widespread adoption.
\item The methodology of this paper is sound, using appropriate quantitative and qualitative techniques.
\end{enumerate}

\subsection{Noteworthy Concerns} %
None.

\end{document}